# The emergence of cooperation from shared goals in the Systemic Sustainability Game of common pool resources

**Authors:** Chengyi Tu[1,2*], Paolo D'Odorico[1], Zhe Li[3], Samir Suweis[4]

**Affiliations:**

[1]Department of Environmental Science, Policy, and Management, University of California, Berkeley; CA 94720-3114, USA.

[2]School of Ecology and Environmental Science, Yunnan University; Kunming, 650091, China.

[3]School of Computer Science and Technology, University of South China; Hengyang, 421001, China

[4]Department of Physics and Astronomy "G. Galilei", INFN, University of Padua; Padova, 35131, Italy.

*Corresponding author. Email: chengyitu@berkeley.edu

**Abstract:**

The sustainable use of common-pool resources (CPRs) is a major environmental governance challenge because of their possible over-exploitation. Research in this field has overlooked the feedback between user decisions and resource dynamics. Here we develop an online game to perform a set of experiments in which users of the same CPR decide on their individual harvesting rates, which in turn depend on the resource dynamics. We show that, if users share common goals, a high level of self-organized cooperation emerges, leading to long-term resource sustainability. Otherwise, selfish/individualistic behaviors lead to resource depletion ("Tragedy of the Commons"). To explain these results, we develop an analytical model of coupled resource-decision dynamics based on optimal control theory and show how this framework reproduces the empirical results.

**One-Sentence Summary:** Understanding sustainable use of common-pool resources through online experiments with shared goals and optimal control theory

Global food and energy security, economic development, and - more generally - the metabolism of human societies strongly depend on natural resources that are under increasing pressure because of the growing demand from human needs and wants (*1*). The sustainable and equitable use of natural resources is a major governance challenge (*2-4*), particularly in the case of resources that are owned and managed by communities as common property systems. This is often the case of common pool resources (CPRs) that are both non-excludable (i.e., their use cannot be easily regulated by limiting or restricting access) and subtractable (one person's use diminishes other persons' ability to use the resource) (*5, 6*). Because of these two properties CPRs are complex to regulate and may be at risk of over-exploitation. Known as "*Tragedy of the Commons*" (*7*), CPR overexploitation leaves future generations empty-handed.

In recent years, however, several studies have demonstrated that in real-world cases users can prevent the *Tragedy of the Commons* and overcome social dilemmas between self-interest and resource overuse by developing institutions of resource governance (*5, 8*). Ostrom (*5*) envisioned that users could self-organize and devise institutions to avoid conditions of tragic overuse. Her work developed a theory to identify effectiveness and limitations of self-governing institutions (*6*), clarifying when and how individuals can overcome the threat of resource exhaustion (*4, 9*). Most of this research, however, concentrated on interactions among users and their impact on decision making (*10*) without accounting for the dynamics of the resource pool and how decisions may be influenced by knowledge on accumulation trends or the severity of resource exhaustion threats. To date, only very few studies have linked decision making to resource dynamics. Indeed, the way such feedback may affect the long-term sustainability of CPRs remains unexplored (*11-13*). Another major challenge in the research on CPR sustainability arises from the lack of real-world experimental data that can be used to test theoretical models of decision-resource dynamics (*14-17*). Overall, we still lack a general testable framework to analyze the sustainability of coupled *Human-Environmental Systems* (HESs) (*18*).

In this study, we develop and run a game, the *Systemic Sustainability Game*, on an online platform designed to collect experimental data that can be used to test theories of HES sustainability (*19*). In particular we test the hypothesis that users may self-organize and engage in collective action conducive to a sustainable governance of CPRs (and associated long-term benefits) by adopting trade-off strategies that balance individual and collective payoff as well as short-term and long-term rewards (*5, 6*).

**Results**

Based on earlier theoretical (*6, 7*), conceptual (*2, 4, 20*) and empirical studies (*5*) on CPRs, we developed a game where $N$ interacting agents, $i = 1, ..., N$, extract resources from the same CPR that at time $t$ exhibits a level, $R(t)$, of available resource (with $0 \le R(t) \le 1$). Each player can harvest the resource pool and obtain an instantaneous payoff proportional to the amount of resource that player has extracted, $U_X = R(t)e_X$, where $e_X$ is the extraction rate and $X$ denotes the user's strategy. We consider two strategies, namely, defection (i.e., $X = D$) and cooperation (i.e., $X = C$), corresponding to a players' decision to extract an amount of resource exceeding or not their sustainable share, respectively. We assume that the CPR grows logistically at rate $T > 0$. If all agents cooperate, then the resource use is sustainable. Conversely, if all agents defect, the resource is over-exploited and eventually exhausted. Thus, the extraction rates (and *rewards*) of defectors are larger than those of cooperators (i.e., $0 < Ne_C < T < Ne_D$) (Materials and Methods, section 1 and Fig S1-5).

We develop an online platform to mimic such HES dynamics and perform experiments involving several interacting players (Fig. 1). At every time step, each player needs to decide whether to cooperate by extracting a sustainable resource fraction or defect by extracting a larger amount. Each player knows: 1) her/his own instantaneous and cumulative payoff (i.e., the total reward accumulated since the beginning of the game); 2) the strategy and payoff of one of the neighbors; 3) the fraction of neighbors that cooperate; 4) the resource level ($R(t)$). $R(t)$ is updated according to a logistic growth minus the extraction (Eq. (1)). The game stops either at time $t = t_f$ — maximum game duration (unknown to the players) — or when the resource is exhausted, whichever happens first. If all players interact with each other, their social network is complete. Alternatively, players are connected only to few other players through a social network with Barabasi-Albert (BA) or small-world (SW) structure (*21*) (Materials and Methods, sections 2 and 3 and Fig. S6-S9).

Each player earns a payoff proportional to the amount of resource he/she has extracted. Participants are divided into two groups playing separately. In type 1 games, participants are rewarded based on their individual performance, while in type 2 games, only participants from the group with the highest collective payoff (the sum of the cumulative payoffs of all group members) receive a reward. The individual reward (in the group that is rewarded) is proportional to the resource extracted by each participant. Thus, type 2 games induce a shared goal at the group level. We run the online experiments at (1) University of California, Berkeley, USA, (2) Yunnan University, China (Fig. S10-S15).

In the absence of a shared goal (type 1 game) we retrieve the *Tragedy of the Commons* (Fig. 2A) because most players tend to maximize their instantaneous benefit (i.e., defect) and exhaust the resource before the end of the maximum game duration. These results are robust with respect to changes in the core variables of the HES such as the number of users ($N$) or the productivity of the system (relative the extraction parameters $e_C$ and $e_D$). When a shared goal is introduced (type 2 game), we find that players do self-organize and reach higher levels of cooperation, thereby allowing for a sustainable use of the resource (Fig. 2B). Ostrom's work (*4*) suggests that such a cooperative behavior is expected to emerge as a result of the existence of common goals within sub-communities of users, as our experiments with shared goals show.

To quantitatively explain these empirical results, we cast the *Systemic Sustainability Game* in a theoretical modeling framework that accounts for the feedback between resource dynamics and human decisions (*11*). Consistent with the experimental setting of the sustainability game, resource dynamics follow a logistic growth with a linear harvest rate accounting for the total extraction rates by all users, which, in turn, depend on their strategies (Eq. (1)). We model strategies as stochastic replicator dynamics (*22*), whereby every player preferentially switches to the strategy of the neighboring player if it leads to a greater payoff.

The mean field equations approximating the full stochastic dynamics of $N$ players harvesting the same resource $0 \leq R(t) \leq 1$ are

$$\frac{dR(t)}{dt} = T\left(R(t)\big(1 - R(t)\big) - R(t)\big(x(t)\hat{e}_C + \big(1 - x(t)\big)\hat{e}_D\big)\right) \quad (1)$$

$$\frac{dx(t)}{dt} = -w(t)R(t)\big(1 - x(t)\big)x(t) \quad (2)$$

where $0 \leq x(t) \leq 1$ is the fraction of players who at time $t$ cooperate, $T > 0$ is the growth rate, $w(t)$ is known as selective pressure, $\hat{e}_C = \dfrac{Ne_C}{T}, \hat{e}_D = \dfrac{Ne_D}{T}$ are normalized extraction parameters of cooperators and defectors with $0 < \hat{e}_C < 1 < \hat{e}_D$.

When players pursue maximization of individual instantaneous payoff the resource is exhausted before the end of the game (*Tragedy of the Commons*). However, if players make their decisions considering the payoff that can be accumulated over all the game duration, thus accounting for resource growth and their ability to harvest it multiple times, then the optimal strategy can be described by a non-trivial optimal control problem:

$$\max_{w(t)} \int_{t_0}^{t_f} TR(t)\big( x(t)\hat{e}_C + \big(1-x(t)\big)\hat{e}_D \big) dt \,, \quad (3)$$

with $R(t)$ and of $x(t)$ given by Eqs. (1)-(2) and where $t_0$ and $t_f$ are the start and end time of the game, respectively. We aim at finding the control variable $w(t)$ and the associated state variables, $R(t), x(t)$, that maximize the average cumulative payoff of players (Eq. (3)) (Materials and Methods, section 4 and Fig. S16-S21).

We use the above optimal control framework to quantitatively interpret the empirical results obtained from the *Systemic Sustainability Game*. We obtain a good fit of the empirical data from the sustainability game (Fig. 3), thus determining the $w(t)$ function that captures the overall strategy followed by players during the game. We use the least squares method to determine $w(t)$ by minimizing the error between the fraction of cooperators observed in the experiments and in the horizon of the optimal control framework. The critical time $\tau_{crt}$ inferred by the fitting procedure corresponds to the time needed by players to reach a self-organized stationary strategy, i.e., $\dfrac{dx(t)}{dt} = 0$ for $t \geq \tau_{crit}$ (Materials and Methods, section 5 and Fig. S22).

Interestingly, we find that in games played in China $\tau_{crt}$ is much smaller in type 2 than in type 1 games, while the opposite is true of experiments carried out in USA (Materials and Methods section 6 and Fig. S23-S25). In other words, in experiments from China cooperation and self-organization are reached sooner under shared goal than individual reward rules, while the opposite is found in the experiments from the USA. Most importantly, in type 2 games (i.e., with shared goals) self-organization leads to a high level of cooperation, thereby allowing for a sustainable resource use. Although this result holds true for both experiments, in China shared goals consistently induced a stronger difference in player decisions and resource levels between the two game types (i.e., with and without shared goals) than in the USA where we observe a relatively high level of cooperation even in type 1 games ($x^* \approx 0.4$ vs $x^* \approx 0.2$ in China). On the other hand, with shared goals players in China self-organize achieving very strong cooperation ($x^* \approx 0.8$), while in the USA we find a 20% increase in the fraction of cooperators ($x^* \approx 0.6$). Nevertheless, in all experiments shared goals (and the presence of a feedback between resource and decision dynamics) are found to be an essential factor for the emergence of cooperation and sustainability in CPRs. As expected, the average cumulative payoff obtained by players is consistently higher in type 2 than in type 1 games, indicating that higher cooperation corresponds to higher average payoff. We also find that the structure of the social network of interactions among players does not significantly affect the outcome of players' decisions. This result, however, may be contributed by the finite size of our experiments (*23*). In fact, extensive numerical simulations of the model (Eqs. (1)-(2)) with larger values of $N$ show that the network structure may affect self-organized cooperation among players ($24$). Specifically, if $w > 0$ networks with relatively small average connectivity tend to sustain higher cooperation with respect to the mean field case (Fig. S6 and S8). Conversely, when $w < 0$ poorly connected social networks tend to be detrimental to cooperation. With high network connectivity the HES dynamics are well captured by the mean field solution (Fig. S7 and S9).

**Discussion**

Understanding the conditions conducive to the sustainable harvesting of CPRs by multiple users remains a major environmental governance challenge. Knowledge of ongoing changes in resource levels, past extraction rates, and associated rewards must play a role in informing decisions on resource extraction. Quite surprisingly, such a resource-decision feedback has remained poorly investigated in the study of CPR dynamics ($12, 13, 16, 17$). Profit maximization by self-interested individuals tends to drive CPRs to exhaustion (*Tragedy of the Commons*) ($7$). Ostrom ($5$) envisioned that users could self-organize and cooperate to avoid conditions of tragic overexploitation and developed a theory explaining the effectiveness and limitations of self-governing institutions. Ultimately, we find that what motivates the emergence of such institutions is the (shared) goal of preservation of the CPR because of the value or "reward" it provides to its users and future generations. Our study demonstrates how shared goals may indeed induce cooperation. Empirical data from our Sustainability Game show that, when players have shared goals ($20$), their behavior optimizes long-term accumulated payoff, leading to self-organized cooperation, while preventing the overexploitation of the CPRs in agreement both with Ostrom's work ($3, 5, 15$) and our modelling framework based on optimal control theory in coupled resource-decision dynamics. We thus prove that common goals favor cooperation and the sustainable harvesting of CPRs. Finally, the empirical data suggest that difference in culture and society may have a relevant impact on players' behaviors since, in absence of shared goals, players from University of California Berkeley self-organize to a higher level of cooperation than Yunnan University. On the other hand, shared goals induce cooperation more effectively in China than in the USA (Fig. S23-S25). These results point to possible effects of culture and related factors on the decision process and outcomes of the sustainability game experiments, a result that will be further investigated through a stand-alone version of the Systemic Sustainability Game (https://systemic-sustainability.com). The open platform will open promising future research paths based on citizenships science experiments ($25$), and allowing for the collection of a large, distributed dataset that can be used to test this and other human-environment system models.

**References and Notes**

1. K. F. Davis, P. D'Odorico, M. C. Rulli, Moderating diets to feed the future. *Earth's Future* **2**, 559-565 (2014).

2. T. Dietz, E. Ostrom, P. C. Stern, The struggle to govern the commons. *Science* **302**, 1907-1912 (2003).

3. E. Ostrom, The challenge of common-pool resources. *Environment: Science and Policy for Sustainable Development* **50**, 8-21 (2008).

4. E. Ostrom, A general framework for analyzing sustainability of social-ecological systems. *Science* **325**, 419-422 (2009).

5. E. Ostrom, *Governing the commons: The evolution of institutions for collective action*. (Cambridge university press, 1990).

6. F. Van Laerhoven, E. Ostrom, Traditions and Trends in the Study of the Commons. *International Journal of the Commons* **1**, 3-28 (2007).

7. G. Hardin, The Tragedy of the Commons. *Science* **162**, 3 (1968).

8. A. Tavoni, M. Schlüter, S. Levin, The survival of the conformist: social pressure and renewable resource management. *Journal of theoretical biology* **299**, 152-161 (2012).

9. O. P. Hauser, D. G. Rand, A. Peysakhovich, M. A. Nowak, Cooperating with the future. *Nature* **511**, 220-223 (2014).

10. C. Hilbe, Š. Šimsa, K. Chatterjee, M. A. Nowak, Evolution of cooperation in stochastic games. *Nature* **559**, 246 (2018).

11. A. R. Tilman, J. B. Plotkin, E. Akçay, Evolutionary games with environmental feedbacks. *Nature communications* **11**, 1-11 (2020).

12. M. A. Janssen, Introducing ecological dynamics into common-pool resource experiments. *Ecology and Society* **15**, (2010).

13. M. Casari, C. R. Plott, Decentralized management of common property resources: experiments with a centuries-old institution. *Journal of Economic Behavior & Organization* **51**, 217-247 (2003).

14. J. M. Anderies *et al.*, The challenge of understanding decisions in experimental studies of common pool resource governance. *Ecological Economics* **70**, 1571-1579 (2011).

15. E. Ostrom, The value-added of laboratory experiments for the study of institutions and common-pool resources. *Journal of Economic Behavior & Organization* **61**, 149-163 (2006).

16. M. Schlüter, C. Pahl-Wostl, Mechanisms of resilience in common-pool resource management systems: an agent-based model of water use in a river basin. *Ecology and Society* **12**, (2007).

17. J.-C. Cardenas, M. Janssen, F. Bousquet, in *Handbook on experimental economics and the environment*. (Edward Elgar Publishing, 2013).

18. B. L. Turner *et al.*, Illustrating the coupled human–environment system for vulnerability analysis: three case studies. *Proceedings of the National Academy of Sciences* **100**, 8080-8085 (2003).

19. O. Amir, D. G. Rand, Economic games on the internet: The effect of $1 stakes. *PloS one* **7**, e31461 (2012).

20. E. Ostrom, *Understanding institutional diversity*. (Princeton university press, 2009).

21. M. Newman, *Networks*. (Oxford university press, 2018).

22. A. Traulsen, J. C. Claussen, C. Hauert, Coevolutionary dynamics: from finite to infinite populations. *Physical review letters* **95**, 238701 (2005).

23. M. A. Nowak, A. Sasaki, C. Taylor, D. Fudenberg, Emergence of cooperation and evolutionary stability in finite populations. *Nature* **428**, 646-650 (2004).

24. C. Tu, S. Suweis, P. D'Odorico, Impact of globalization on the resilience and sustainability of natural resources. *Nature Sustainability* **2**, 283-289 (2019).

25. E. Awad *et al.*, The moral machine experiment. *Nature* **563**, 59-64 (2018).

**Acknowledgments:** Acknowledgments follow the references and notes list but are not numbered. Start with text that acknowledges non-author contributions and then complete each of the sections below as separate paragraphs.

**Funding:** This work was funded in part by Microsoft AI for Earth; Experimental Social Science Laboratory (Xlab) at University of California Berkeley; Yunnan University project C176210103. SS acknowledges *LINCON* INFN and BIRD_UNIPD grants.

**Author contributions:**

Conceptualization: CT, PD, SS

Methodology: CT, ZL, PD, SS

Investigation: CT, PD, SS

Visualization: CT, SS

Funding acquisition: CT, PD, SS

Project administration: PD, SS

Supervision: PD, SS

Writing – original draft: SS

Writing – review & editing: CT, PD, SS

**Competing interests:** Authors declare that they have no competing interests.

**Data and materials availability:** Data and code for all analyses are available at OSF (https://osf.io/txus6/).

## Supplementary Materials

Materials and Methods

Figs. S1 to S25

Tables S1

References (*1–12*)

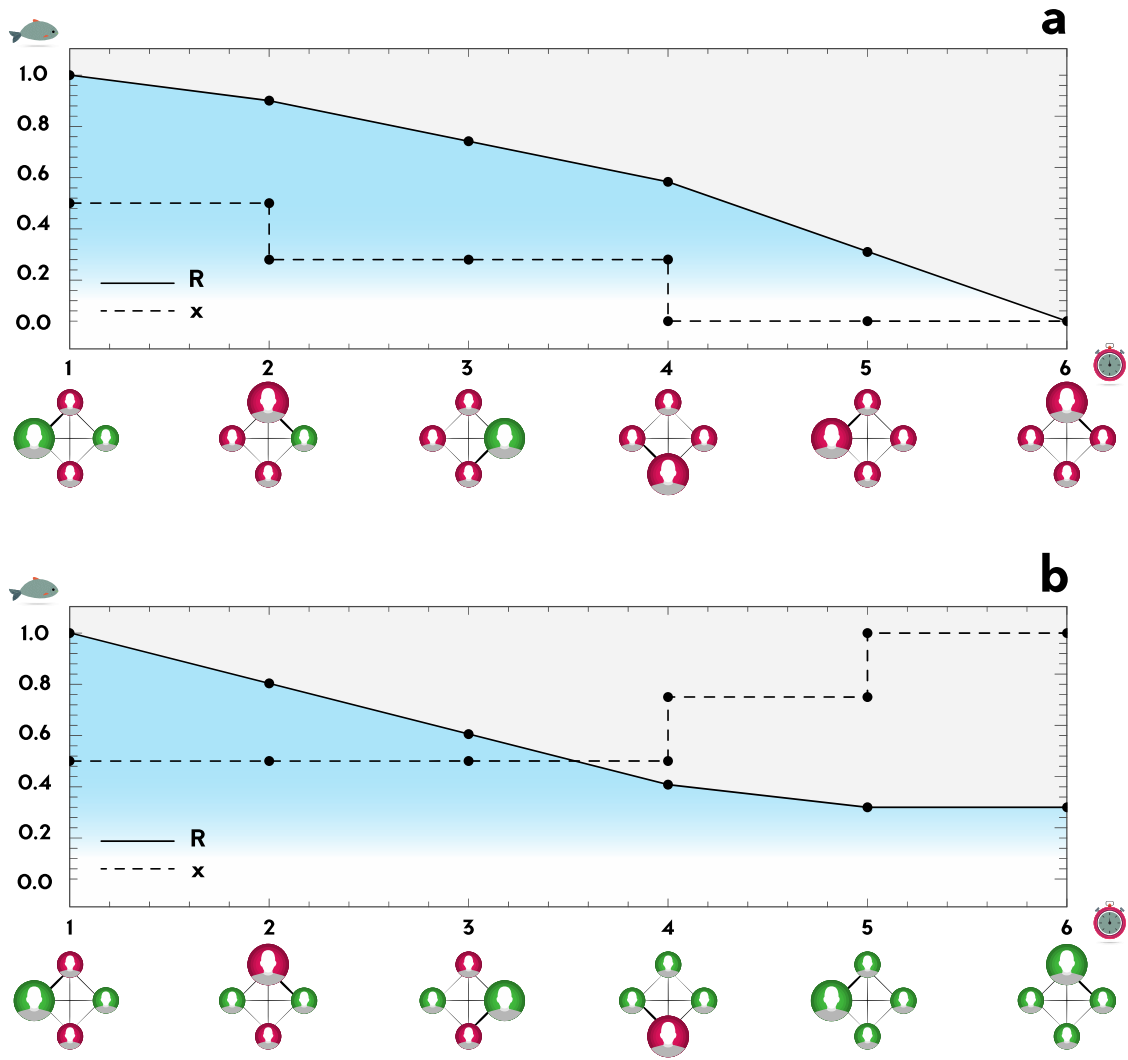

**Fig. 1. An illustration of the *Systemic Sustainability Game* for case of 4 players.** At each step, the active player (the highlighted node) compares her/his payoff to one of her/his neighbors' where cooperators are shown in green and x denotes the fraction of players who cooperate; defectors are shown in red. The user's resource extraction depends on the chosen strategy. Defectors extract more than cooperators thus having a higher instantaneous reward. (**a**) If most players defect, then resource, *R*, is not sustainably harvested, and the reward tends to zero. (**b**) If most players cooperate to achieve collective benefits, then resource can be sustainably harvested over longer time frames and can provide an overall higher cumulative reward (i.e., the total reward accumulated since the beginning of the game).

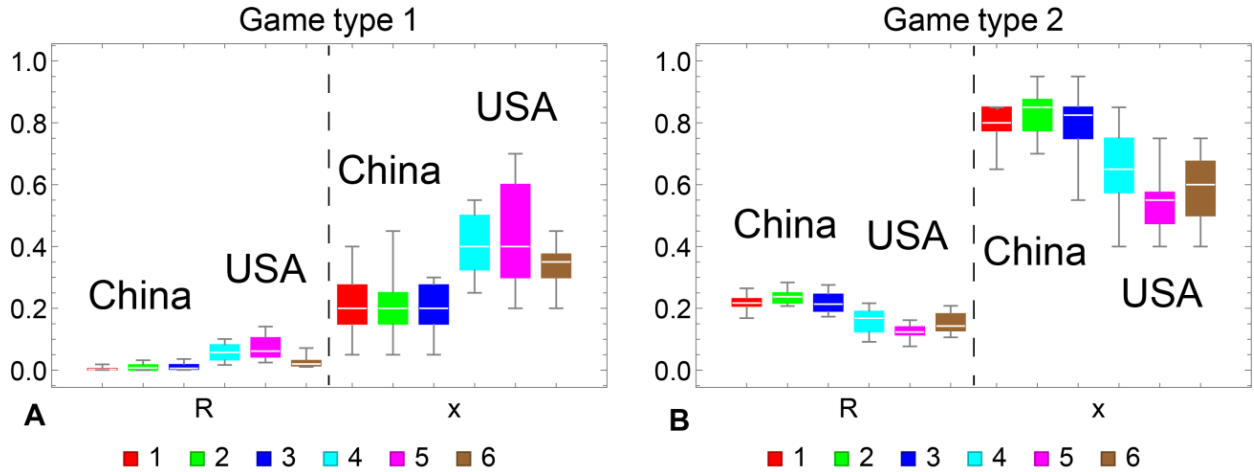

**Fig. 2**. **The comparison between stable equilibria reached with the two game types.** (**A**) game type 1 (no shared goals) and (**B**) game type 2 (with shared goals) with different social network topologies with players from China or the USA. All experiments used the same growth rate $T = 2$, normalized extraction parameter $\hat{e}_C = 0.7, \hat{e}_D = 1.1$, initial condition $R_0 = 0.5, x_0 = 0.5$ and max game duration $t_f = 40$. Experiments 1-3 were performed at the Yunnan University, China; experiments 4-6 were performed at the University of California, Berkeley, USA.

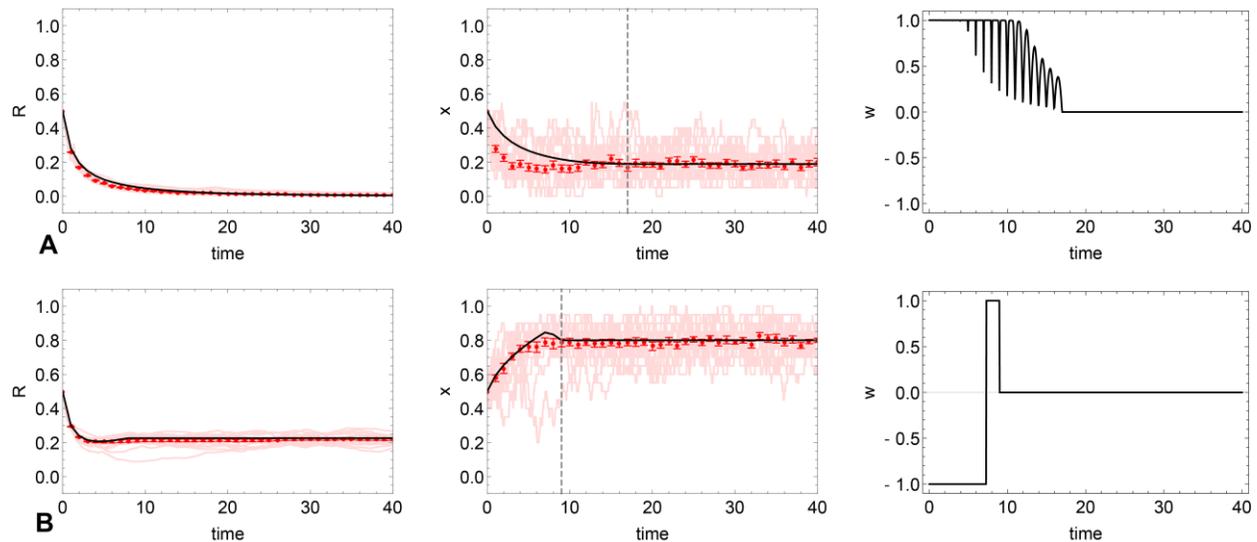

**Fig. 3**. **Comparison between experimental results of the *Systemic Sustainability Game* run in China with complete network and its theoretical prediction.** (**A**) Game type 1: participants are rewarded based on their individual performance (**B**) Game type 2: participants are divided into two groups (N=20 participants each) and the rewards is paid only to the group that has obtained the highest cumulative reward. Parameter values were the same as in Fig. 2. The associated theoretical prediction is determined through our optimal control framework that allows for the direct inference of $w(t)$ from the empirical data. Light red lines represent single realizations of the experiment. The red points and bars are the corresponding averages and standard errors at each integer time. The black lines in the left and central panels represent theoretical prediction. The black dashed lines represent critical time.

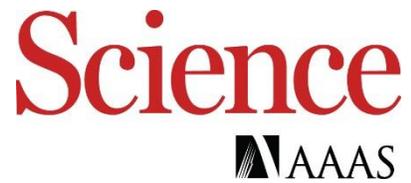

Supplementary Materials for

The emergence of cooperation from shared goals in the Systemic Sustainability Game of common pool resources

Chengyi Tu, Paolo D'Odorico, Zhe Li, Samir Suweis.

Correspondence to: chengyitu@berkeley.edu

**This PDF file includes:**

> Materials and Methods
> Figs. S1 to S25
> Tables S1

**Materials and Methods**

# 1 Theoretical derivation

### 1.1 Modelling Environmental Feedback: Resource Dynamics

We start from the simplest case of one resource pool with time dependent volume $R(t)$ and $N$ players extracting value from it. Each player can select between with two strategies: extracting a sustainable amount of resource (cooperation), or extracting a larger, unsustainable amount (defection). We denote the fraction of cooperators in the system with $x$ (and thus defectors are $1-x$). The evolution of the resource is described by a logistic function, $TR(t)\left(1-\dfrac{R(t)}{K}\right)$ where $T > 0$ is the natural growth rate, while $K$ is the carrying capacity. The extraction rates adopted by individual players selecting cooperation and defection are $e_C, e_D$, respectively, with $0 < Ne_C < T < Ne_D$. Therefore, the total amount of resources extracted by all players is $E = N_C e_C + N_D e_D = N\left(xe_C + (1-x)e_D\right)$.

We can thus write the differential equation describing the evolution of the resource volume $R(t)$ as

$$\frac{dR(t)}{dt} = TR(t)\left(1-\frac{R(t)}{K}\right) - R(t)E = TR(t)\left(1-\frac{R(t)}{K}\right) - NR(t)\left(x(t)e_C + \left(1-x(t)\right)e_D\right). \quad \text{(S1)}$$

### 1.2 Evolutionary dynamics of user's strategies

The dynamics of the players' strategies are expressed by a stochastic replicator rule, i.e., a player switches strategy depending on the payoff differences with one of the neighbors. We will start by describing the stochastic microscopic process through a master equation and then we will take the continuum large $N$ limit and formulate the dynamics in terms of a Fokker-Planck (forward Kolmogorov) equation.

#### 1.2.1 Microscopic Stochastic Dynamics

Let $i = 1 \ldots N$ label the player (node) in a given social interaction network $M$ and let us call $U_i$ the payoff of player $i$. For each player, the payoff at any given time $t$ will be proportional to the amount of extracted resources by that player, i.e. $U_C = R(t)e_C$ and $U_D = R(t)e_D$ for cooperators and defectors, respectively. We highlight that the efforts $e_C, e_D$ are fixed at the beginning, while the payoffs $U_C, U_D$ vary in time depending on the available resource volume. Nevertheless, at each time step, $U_C < U_D$ as the extraction rate of defectors is larger than the one of cooperators.

Following the replicator rule, the players' strategy is modeled by: (1) randomly selecting one player $i$; (2) randomly selecting one of $i$'s neighbors, $j$; (3) allowing the strategy of player $i$ to switch to the strategy of player $j$ with probability $p = \dfrac{1}{2} + \dfrac{w}{2}\dfrac{U_j - U_i}{\Delta U_{max}}$ where $U_i, U_j$ are the payoffs of players $i$ and $j$ (If player $i$ selects cooperation, $U_i = U_C$; if $i$ selects defection, $U_i = U_D$ ), $\Delta U_{max}$ is the maximum possible payoff difference for guaranteeing that $0 \leq p \leq 1$ ( $\Delta U_{max} = \max |U_D - U_C| = e_D - e_C$ when $R = 1$ ) and $w$ is the selection pressure (*1, 2*). In this setting, the probabilities to switch strategy from defection to cooperation, and vice versa are $p^+ = \dfrac{1}{2} + \dfrac{w}{2} \dfrac{Re_C - Re_D}{e_D - e_C} = \dfrac{1}{2} - \dfrac{w}{2} R$ and $p^- = \dfrac{1}{2} + \dfrac{w}{2} \dfrac{Re_D - Re_C}{e_D - e_C} = \dfrac{1}{2} + \dfrac{w}{2} R$ , respectively, with $-1 \leq w \leq 1$ . In the classic game theory framework $w(t)$ is a positive constant for players to pursue maximization of individual *instantaneous* payoff at a given time $t$ . With positive $w(t)$ the probability that a player selects defection is always larger than the probability of adopting cooperation, and the inevitable destiny of the common pool resource $R(t)$ is the complete exhaustion. Defection, however, does not necessarily maximize the cumulative payoff over time. Indeed, higher cumulative payoffs can be obtained in the presence of some cooperation (i.e., periods with $w(t) < 0$ ).

The corresponding transition probability at time $t$ of passing from $N_C$ cooperators to $N_C \pm 1$ given an amount of resource $R(t)$ and selection pressure $w(t)$ is

$$T^{\pm}(N_C \mid R, w; t) = p^{\pm} \frac{N_C}{N} \frac{N - N_C}{N} = \left( \frac{1}{2} \mp \frac{w}{2} R \right) \left( \frac{N_C}{N} \frac{N - N_C}{N} \right).$$

If players interact through a complete network (i.e., each node is connected to all other nodes), then the Master Equation for the evolution of the probability $P^{\tau}(N_C)$ of having $N_C$ cooperators at time $\tau$ is

$$\begin{aligned} P^{\tau+1}(N_C) - P^{\tau}(N_C) = \ & P^{\tau}(N_C - 1)T^+(N_C - 1 \mid R, w; \tau) + P^{\tau}(N_C + 1)T^-(N_C + 1 \mid R, w; \tau) \\ & - P^{\tau}(N_C)T^-(N_C \mid R, w; \tau) - P^{\tau}(N_C)T^+(N_C \mid R, w; \tau) \end{aligned}.$$

### 1.2.2 Continuous approximation

Adopting classic evolutionary game framework, we can derive a general Fokker-Planck equation for the probability density of having a fraction $x$ of players using a cooperative strategy, $\rho(x)$ .We start by the above Master Equation and substituting $x = \dfrac{N_C}{N}, t = \dfrac{\tau}{N}$ , and $\rho(x; t) = N P^{\tau}(N_C)$ yields to

$$\begin{aligned} \rho(x; t + N^{-1}) - \rho(x; t) = \ & \rho(x - N^{-1}; t)T^+(x - N^{-1} \mid R, w; t) + \rho(x + N^{-1}; t)T^-(x + N^{-1} \mid R, w; t) \\ & - \rho(x; t)T^-(x \mid R, w; t) - \rho(x; t)T^+(x \mid R, w; t) \end{aligned}.$$

For $N$ much larger than 1, the probability densities and the transition probabilities are expanded in a Taylor series around $x(t)$ . Neglecting higher order terms in $N^{-1}$ , we obtain the Fokker-Planck equation

$$\frac{d}{dt}\rho(x; t) = -\frac{d}{dx}[a(x \mid R, w; t)\rho(x; t)] + \frac{1}{2}\frac{d^2}{dx^2}[b^2(x \mid R, w; t)\rho(x; t)] \quad \text{(S2)}$$

where $\qquad a(x \mid R, w; t) = T^+(x \mid R, w; t) - T^-(x \mid R, w; t)$ and

$b(x \mid R, w; t) = \sqrt{\dfrac{1}{N}\left(T^+(x \mid R, w; t) + T^-(x \mid R, w; t)\right)}$ . Since subsequent time steps are independent, the noise is not correlated in time and the Ito calculus can be applied to derive the corresponding Langevin equation:

$$\frac{dx(t)}{dt} = a(x \mid R, w; t) + b(x \mid R, w; t)\zeta \quad \text{(S3)}$$

where $\zeta$ is an uncorrelated Gaussian white noise. For $N \to \infty$, the diffusion term $b(x \mid R, w; t)$ vanishes with $1/\sqrt{N}$ and a deterministic equation is obtained. Therefore, the mean field equation for the evolution of the players strategy is

$$\frac{dx(t)}{dt} = a(x \mid R, w; t) = x(t)\big(1 - x(t)\big)\big(p^+(t) - p^-(t)\big)$$

and by substituting $p^+(t) = \frac{1}{2} - \frac{w(t)}{2}R(t)$ and $p^-(t) = \frac{1}{2} + \frac{w(t)}{2}R(t)$, we obtain

$$\frac{dx(t)}{dt} = -w(t)R(t)x(t)\big(1 - x(t)\big). \quad \text{(S4)}$$

### 1.3 Human-Environmental System coupled equations

We can now write the full dynamics of the human-environmental system (HES) coupled equations as

$$\begin{cases} \dfrac{dR(t)}{dt} = TR(t)\left(1 - \dfrac{R(t)}{K}\right) - NR(t)\big(x(t)e_C + \big(1 - x(t)\big)e_D\big) \\[2mm] \dfrac{dx(t)}{dt} = -w(t)R(t)x(t)\big(1 - x(t)\big) \end{cases} \quad \text{(S5)}$$

This macroscopic equation can also be considered as a mean field game with environmental feedback (*3*). For simplification, we set $K = 1$ by normalizing the resource volume $R$ between 0 and 1. Defining $\hat{e}_C = \dfrac{Ne_C}{T}, \hat{e}_D = \dfrac{Ne_D}{T}$, the Eq. (S5) reads as

$$\begin{cases} \dfrac{dR(t)}{dt} = T\big(R(t)\big(1 - R(t)\big) - R(t)\big(x(t)\hat{e}_C + \big(1 - x(t)\big)\hat{e}_D\big)\big) \\[2mm] \dfrac{dx(t)}{dt} = -w(t)R(t)x(t)\big(1 - x(t)\big) \end{cases} \quad \text{(S6)}$$

where $T > 0$, $0 < \hat{e}_C < 1 < \hat{e}_D$, $0 \le x \le 1$, $0 \le R \le 1$. The equilibrium and stability of this coupled differential equations do not depend on $T$. In the following analysis, otherwise made explicit, we set $T = 2$ without loss of generalization.

### 1.3.1 Individual based model

We also simulate the corresponding stochastic individual based model where the players' strategies are updated through the stochastic replicator rule presented above, while $R(t)$ is updated through a deterministic logistic map and users resource extractions. The microscopic dynamics after initialization consist of the following steps:

(1) select one player, $i$, randomly;

(2) select one of $i$'s neighbors, $j$, randomly;

(3) allow for the strategy of player $i$ to switch to the strategy of player $j$ with probability

$$p = \frac{1}{2} + \frac{w}{2}\frac{U_j - U_i}{\Delta U_{max}};$$

(4) update $x[k] = \dfrac{N_C}{N}$ where $k$ is the discrete time, $N_C$ is the number of players selecting cooperation and $N$ is the number of players; then the resource dynamics can be expressed as

$$R[k] = R[k-1] + \frac{1}{N}\Big(TR[k-1]\big(1-R[k-1]\big) - NR[k-1]\big(x[k-1]e_C + \big(1-x[k-1]\big)e_D\big)\Big) \ ,$$ which is simply the discretization of Eq. (S6).

### 1.3.2 Comparison between individual based model and mean field HES

If the network size is small, i.e., the number of players is relatively small, stochasticity in the microscopic dynamics may be so strong to force the system away from the expected stable equilibrium of the HES. However, if we average over several different stochastic realizations, we can see that the mean field approach well describes the averaged stochastic simulations. As network size increases, stochasticity plays a less relevant role and each stochastic realization tends to the HES coupled equations as shown on Fig. S1.

### *1.4 Equilibrium and attractors*

### 1.4.1 Resource dynamics with fixed strategy

If all players adopt the same strategy $X$ with extraction rate $e_X$, then the single resource dynamics become $\dfrac{dR(t)}{dt} = T\big(R(t)\big(1-R(t)\big) - \hat{e}_X R(t)\big)$ where $\hat{e}_X = \dfrac{e_X N}{T}$. Its trajectory is $R(t) = \dfrac{R_0(\hat{e}_X - 1)}{\hat{e}_X^{(T(\hat{e}_X - 1)}(R_0 + \hat{e}_X - 1) - R_0}$ where $R_0$ is the initial condition. The collective cumulative payoff of all players from time $0$ to $t_f$ is $\mathcal{Q}(t_f) = \int_0^{t_f} TR\hat{e}_X dt = \int_0^{t_f} Re_X N dt$. Therefore, the cumulative payoff of each player (individual, on average) from time $0$ to $t_f$ is

$$\mathcal{P}(t_f) = \frac{1}{N}\mathcal{Q}(t_f) = \int_0^{t_f} Re_X dt \qquad\qquad , \qquad\qquad \text{with}$$

$\mathcal{Q}(t_f) = \int_0^{t_f} T\dfrac{R_0(\hat{e}_X - 1)}{\hat{e}_X^{(T(\hat{e}_X - 1)}(R_0 + \hat{e}_X - 1) - R_0}\hat{e}_X dt = \hat{e}_X\left(\log\big(R_0 - (\hat{e}_X + R_0 - 1)e^{Tt_f(\hat{e}_X - 1)}\big) - Tt_f(\hat{e}_X - 1) - \log\big(1-\hat{e}_X\big)\right)$ .

In particular, we compare the two extreme scenarios, $x^* = 0$ (all players defect) and $x^* = 1$ (all players cooperate), for which we found that the cumulative payoff of each player is

$$\mathcal{P}(t_f) = \frac{1}{N}\hat{e}_X\left(\log\big(R_0 - (\hat{e}_X + R_0 - 1)e^{Tt_f(\hat{e}_X - 1)}\big) - Tt_f(\hat{e}_X - 1) - \log\big(1-\hat{e}_X\big)\right) \text{ with } X = C \text{ if } x^* = 1,$$

and $X = D$ if $x^* = 0$, and $R_0 = R(0)$. In the case of short-term horizons (e.g., $t_f < 1.5$), the most rewarding strategy is attained with defection (Fig.S2 A). Conversely, cooperation and the associated sustainable resource use allow for higher cumulative payoffs in the long term if the game lasts a sufficiently long time (e.g., $t_f > 1.5$).

If we further consider a constant fraction of cooperators and defectors, and $x^*$ is the (constant) fraction of cooperators, then the accumulated payoff up to a time $t_f$ is given by

$$\mathcal{P}(t_f) = \frac{1}{N} \int_0^{t_f} TR(x^* \hat{e}_C + (1-x^*)\hat{e}_D)dt = \int_0^{t_f} R(x^* e_C + (1-x^*)e_D)dt , \quad \text{(S7)}$$

and can be solved analytically, by substituting the time dependent solution of $R(t)$. In particular, solving the resource equation at stationarity for $x^*$ constant at intermediate values between 0 and 1 ( $0 < x^* < 1$ ), the stable solution is $R^* = 0$ if $0 < x^* \le x_l$ where $x_l = \dfrac{-1 + \hat{e}_D}{-\hat{e}_C + \hat{e}_D}$ . Therefore, if $0 < x^* \le x_l$ at steady state the instantaneous payoff is 0 for all players, no matter of whether they cooperate or defect. Conversely, if $x_l < x^* < 1$ , the stable resource level is $R^* = (1 + \hat{e}_D(-1 + x^*) - \hat{e}_C x^*)$ and players earn a payoff because $R^* > 0$ , as shown in Fig. S2 B. We thus find that the most rewarding strategies, if the game is long enough, are attained when $x_l < x^* < 1$ . Further, when $x_l < x \le x_l + 1/N$ , one cooperator changes to defector leads to his instantaneous payoff decreases from $R(x)e_C$ to 0, while one defector changes to cooperator leads to his instantaneous payoff decreases from $R(x)e_D$ to $R(x + 1/N)e_C$ . In other words, in this situation one player cannot increase his own expected payoff by changing his strategy while the other players keep unchanged, so any feasible $x$ satisfying $x_l < x \le x_l + 1/N$ is a stable evolutionary strategy. In particular, because player number $N$ is an integer, there is only one $x$ that holds condition $x_l < x \le x_l + 1/N$ for any given normalized extraction parameters $\hat{e}_C, \hat{e}_D$ and we call it Nash equilibrium $x_{Nash}$ .

### 1.4.2 Strategies evolutionary dynamics with fixed resources

If the resource volume is constant, then there is no environmental feedback, and the stationary states of strategy dynamics do not depend on the resources. In fact, the equation $\dfrac{dx(t)}{dt} = -wRx(t)\big(1-x(t)\big)$ has two equilibria $x^* = 0$ and $x^* = 1$ . When $w > 0$ , the stable equilibrium is $x^* = 0$ (i.e., all players in the long term will defect); When $w < 0$ , the stable equilibrium is $x^* = 1$ (i.e., all players in the long term will cooperate (Fig. S2 C)). In other words, when environmental feedback is absent, we have two trivial stationary solutions depending on the sign of $w$ . More interestingly, if resources are constant, then we can find the transient solution of the above equation, i.e., $x(t) \to -\dfrac{x_0}{e^{Rtw}x_0 - e^{Rtw} - x_0}$ , where $x_0$ is the fraction of cooperators at initial time $t = 0$ . Additionally, when $w = 0$ , $x^* = x_0$ . In Fig. S3 we show that the analytical solution correctly describes the numerical simulation of the evolution of the fraction of cooperators.

### 1.4.3 Full coupled Human-Environmental System dynamics

We cannot find the analytical temporal solution of the HES given by Eq. (S6), but we can study its stationary solutions and their stability. The Jacobian matrix describing the linearized dynamics around a given stationary solution $\{R^*, x^*\}$ is

$$J(R^*, x^*) = \begin{pmatrix} T\Big[(1 - 2R^*) - \hat{e}_D(1 - x^*) - \hat{e}_C x^*\Big] & (\hat{e}_D - \hat{e}_C)R^*T \\ -wx^*(1 - x^*) & -wR^*\Big[(1 - x^*) - x^*\Big] \end{pmatrix} \quad \text{(S8)}$$

while the stationary solutions of Eq. (S6) are: $\mathbf{s}_1 = \{R = 0, \forall x\}$ and $\mathbf{s}_2 = \{R = 1 - \hat{e}_C, x = 1\}$.

The corresponding HES eigenvalues evaluated from Eq. (S8) for the above stationary solutions are: $\boldsymbol{\lambda}_1 = \{0, T\left(1 - \hat{e}_D(1 - x^*) - \hat{e}_C x^*\right)\}$ and $\boldsymbol{\lambda}_2 = \{T(\hat{e}_C - 1), w(1 - \hat{e}_C)\}$. Therefore, based on the Hartman–Grobman theorem (4) we cannot state anything about the stability of $\mathbf{s}_1$ as the corresponding Jacobian has a zero eigenvalues. However, in this case we observe that system is neutrally stable, i.e., while $R$ at stationarity is always is zero and return to zero even if system is perturbed, there is not a stable stationary fraction of cooperators $x^*$, rather it changes every time the system is perturbed. Assuming that $R(t)$ is a fast variable with respect to $x(t)$, then we can assume that when $x(t) \rightarrow x^*$, then $R(t)$ is small and we can neglect the quadratic term of upper part of Eq. (S6). In this approximation, we have that $R(t) \approx e^{tT - \hat{e}_D tT - \hat{e}_C tTx^* + \hat{e}_D tTx^*} R(0)$ and substituting it into the lower part of Eq. (S6), we can find the self-consistency equation

$$x^* = \frac{1}{1 - e^{\frac{(\hat{e}_C - \hat{e}_D)R(0)w}{T(1 + \hat{e}_D(-1 + x^*) - \hat{e}_C x^*)}}(-1 + x(0))},$$

which shows that the players stationary strategy $x^*$ depends on both the initial condition of $x$ and $R$ (see Fig. S4), i.e., the steady resource level reaches zero at stationarity, but the strategy evolutionary dynamics is only neutrally stable. Therefore, all the solutions in $\mathbf{s}_1$ are neutrally stable, i.e., while $R$ at stationarity is always zero and return to zero even if the system is perturbed, there is not a stable stationary fraction of cooperators $x^*$, as it changes every time the system is perturbed. On the other hand, $\mathbf{s}_2$ is an unstable equilibrium for $w(t) > 0$, while it is stable for $w(t) < 0$.

For $w = 0$, $x = x_0$ is constant so that we can solve the resource dynamics given by upper part of Eq. 6 and the stationary solution is given by $R = \min\{0, 1 - \hat{e}_D - \hat{e}_C x_0 + \hat{e}_D x_0\}$ and the sustainability of the resource depends on the constant fraction of cooperator, i.e., resource is depleted if $x_0 \leq \frac{\hat{e}_D - 1}{\hat{e}_D - \hat{e}_C}$. In summary, as shown by Fig. S5, for $w > 0$ the only (neutrally) stable state is $\mathbf{s}_1$, while for $w < 0$ the only stable solution is $\mathbf{s}_2$.

## 2 Effect of the social network structure on the dynamics

Up to now we have neglected the effect of the social network structure among players, i.e., we have considered that all players may interact with all the other players in the game. Typically, however, players do not have such information, rather they are connected with only few other players and there exist a structure in the network through which players interact. To test the effect of such a network structure on the resource sustainability and strategy evolution, we perform numerical simulations, where only neighborhood players extracted from a given network compare their payoff, i.e. we simulate spatially explicit dynamics.

Moreover, we want to test the effect of stochastic fluctuations due to a number of agents playing the game. In fact, if the number of players is small, stochasticity may be so strong that forces the system far from the expected stable equilibrium obtained by the mean field approach. In this case, variability in the fate of the resource and in the evolution of the strategy is very high and no prediction can be made. Substituting $T^{\pm}$ to the definition of the diffusion parameter b shown in Eq. (S2), i.e., $b(x\,|\,R,w;t) = \sqrt{\dfrac{1}{N}\big(T^{+}(x\,|\,R,w;t) + T^{-}(x\,|\,R,w;t)\big)}$ , we get $\max[b] = \sqrt{\dfrac{1}{4N}}$ when $N_C = \dfrac{N}{2}$. If $b$ is sufficiently large, stochasticity is so strong that the system is driven far from the expected stable configuration.

As shown in Fig. S6 and S7 although single stochastic trajectories may deviate from the mean field solution, in general if we average over many realizations both resource and strategy dynamics are quite close to the ones obtained neglecting spatial network effects and stochasticity. We highlight that for classic replicator dynamics with environmental feedback (case $w = 1$), network structures with low connectivity (e.g. rings and trees) increase the level of cooperation and lead to a greater steady state resource level. On the other hand, if we challenge cooperative dynamics ($w = -1$) the same network structures tend to decrease the cooperation among players.

If we increase the number of players and repeat the same simulations, then the effect of intrinsic demographic noise is reduced, and the structure of the social networks has a more relevant effect on the dynamics. As shown in Fig. S8 and S9 with the same network structures, not only does the agreement between average trajectories over many realizations and theoretical prediction improves, but also the agreement between each realization and theoretical prediction. Particularly, the average trajectory calculated over many realizations is almost the same as the theoretical prediction for complete, BA and SW networks, while the spatial effects of the stochastic dynamics are noticeable in the case of low-connectivity networks such as chain networks and the tree networks. This confirms the fundamental role played by connectivity on the level of cooperation at steady state. Such effects should be further investigated in future work.

## 3 Experimental design

Our experiment is run at University of California, Berkeley, USA and Yunnan University, China. It was approved by the University of California, Berkeley Committee on the Use of Human Subjects, and informed consent was obtained from all subjects prior to beginning the experiment. Participants in each university were recruited online for research purposes. Participants registered by providing their email address. Once registered, they were placed into a recruitment pool for future experiments. Participants were then sent an announcement with a link to participate in an experiment.

### 3.1 Experiment procedure

We recruited more than 40 participants and divided them into 2 groups for each experiment. Each group has 20 player seats that are grabbed by 20 participants. One experiment includes two game types and two groups play the same game type in parallel. Game type 1: The participants received a monetary reward on the basis of their individual performance; Game type 2: The participants received a monetary reward on the basis of their performance as a group. Specifically, only the participants in the group that earns the highest group payoff, (i.e., which is the sum of the payoff earned by all participants in that group), received a reward. Within that group, participants were rewarded on the basis of their individual performance (the same as game type 1). The experimental procedure is as follows:

(1) Experimental Social Science Laboratory (Xlab) at University of California Berkeley issued an announcement to its volunteer pool. The volunteers who are interested in this experiment registered on the link http://13.70.2.226:9528/, and read the instructions.

(2) The study started with a Zoom® session. The link to the Zoom session was emailed and remained available on the Xlab's management toolbox for 10 minutes before the start of the experiment. The participants were assigned to one of two Zoom chats corresponding to two different groups. Each group had the same size and harvests a separate common-pool resource.

(3) Each group first received one link to play 10 games of type 1.

(4) Subsequently, each group received another link to play 10 games of type 2.

## 3.2 Experiment announcement

An example of announcement is shown as following.

*You are invited to participate in a study evaluating the effects of individual decisions on the use of shared resources, conducted by Prof. D'Odorico and Dr. Tu at the University of California, Berkeley, Department of Environmental Science, Policy, and Management. The experiment will use an online interactive platform to play some games. It will last about 45 minutes.*

*The study will start with a Zoom session. The link to the Zoom session will be emailed and will be available on Sona 10 minutes before the start of the experiment. You will be assigned to one of two Zoom chats corresponding to two different groups. Each group has the same size and harvests a separate common-pool resource. Each group will first receive one link to play 10 games with game type 1 and then receive another link to play 10 games with game type 2.*

*Game type 1: As one of the participants in the experiment, you will receive a monetary reward on the basis of your individual performance.*

*Game type 2: As one of the participants in the experiment, you will receive a monetary reward on the basis of your performance as a group. Specifically, only the participants in the group that earns the highest group payoff, sum of the payoff earned by all individuals in that group, will receive a reward. Within that group, participants will be rewarded on the basis of their individual performance (the same as game type 1), while players in the other group will not receive any reward.*

*At the end of one game, each player is returned to the main interface and should select the next game until all games have been played. If you encounter any problems during experiment, please refresh the web-page.*

*After signing up, register, then read the instructions.*

## 3.3 Parameter configuration and initial condition

We set size $N = 20$, growth rate $T = 2$, normalized extraction parameter $\hat{e}_C = 0.7, \hat{e}_D = 1.1$, initial condition $R_0 = 0.5, x_0 = 0.5$ and game time $t_f = 40$. Each game type has 10 games with same setting for both groups. At last, we have 20 realizations for the both game types respectively.

## 3.4 Architecture of the interactive platform

The interactive platform decouples front-end and back-end by Vue.js (*5*) and Node.js (*6*). Front-end web-page and back-end server exchange data by standard JSON method (*7*). Before the experiment starts, the manager inputs initialization information by management tool, then the back-end server reads them and initializes all games in memory. After the experiment starts, user logins the front-end web-page by submitting their account's login information. If the back-end verifies them, it adds a Cookie to the response as a credential of following exchange. At this point the user becomes a valid player and then enters the main page a link to the available games appear. When player enters one game, he/she is assigned to one network node if he/she is a new player, and then front-end requests model state from back-end for initializing game page. After the player makes one decision in the game page, the control layer of the back-end receives a request from front-end, and then sends it to the model layer. The model layer changes the model state, and then responds to the new state to the control layer, finally the control layer responds providing the required information to the front-end. At the same time, the control layer writes one record in the database, including information on the state variables. After one game reaches a given number of steps, the game ends and the player returns to main page to select a new game until all games have been played. When all games are terminated, the experiment ends. At this point, the manager can export the data. The software flow chart and recorded variables are shown in Fig. S10 and Table S1.

### 3.5 Operational guide for the interactive platform

In this section, we introduce how to use our interactive platform titled "Sustainability Game".
*Step 1. Register/Login.* (see Fig. S11).
*Step 2 Instructions.* After registering, participants are asked to read the following instructions (see Fig. S12):

*"You are now participating to an interesting experiment that includes many different games. Please read the following instructions carefully. A good understanding of these instructions will allow you to perform better in the experiment and receive a better payoff.*

*You will play a game in which you act as a member of a community that relies on a common-pool resource (for instance a group of herders using the same pastures or a fishing community relying on the same fisheries). Every player can harvest/extract resources from the common pool. Each player needs to decide on the amount of resource he/she wants to extract by choosing between two options (or "strategies"), namely, "cooperation" and "defection". The payoff of each player will be proportional to the amount of resources that player has extracted during the game. Each game will entail several time steps and at every round you will have to choose between these two possible strategies (i.e., cooperation and defection). If you select defection, you extract a greater amount of resources than what it is (on average) sustainable. This means that if the other players do the same, then the resource will be exhausted/depleted. Conversely, if you select cooperation, you extract a smaller amount of resource. In this case the resource will not be exhausted, provided that the other players do the same (i.e., cooperate). If in the course of the game the resources are completely depleted, the game stops and the players will not be able to earn additional payoff. During each game you have the following information: (1) your strategy, (2) your payoff, (3) the strategy of one of your neighbors, (4) the payoff of this neighbor, (5) the resource volume and (6) in some games the fraction of your neighbors that is cooperating. The resource volume is expressed by a number ranging between 0 and 1. If it is 0, it means that all resources have been exhausted.*

*For instance, imagine that you are a fisherman catching fish from an open-access lake. If the lake is overfished, the fish stock will be exhausted and no fish will be left for future users. You will have the following information: your strategy, your fish catch, the strategy of one of your neighbors, his/her fish catch and the amount fish that is now left in the lake. Then you will have to select the strategy for the next time step.*

*As one of the participants in the experiment, you will receive a monetary reward. Game type 1: The participants will be rewarded on the basis of their individual performance. For example,*

each group has 20 player seats and average reward is 10$. More than 20 participants (We recruited more participants than available "seats" to ensure that all seats are filled) grab fixed 20*10$=200$ according to their individual payoff. The rate between payoff value and real-world money is not fixed, for example payoff value 1 = 0.01$. During each game, for a higher reward one participant should obtain a higher payoff than others rather than all participants obtain similarly high payoff; Game type 2: The participants will be rewarded on the basis of their performance as a group. Specifically, only the participants in the group that earns the highest group payoff, sum of the payoff earned by all individuals in that group, will receive a reward. Within that group, participants will be rewarded on the basis of their individual performance (the same as game type 1). For example, if on average each player earns 10$ and each group has 20 player seats, then the group with the highest total payoff will receive 200$ and each player in that group will be paid proportionally to her/his individual performance, while players in the other group will not receive any reward. At every round you shall select your strategy only based on the information appearing on the screen. Enjoy the game!"

*Step 3 Game Selection.* As soon as the experiment starts one game is available on the screen (see Fig. S13). The following message appears on the screen:

Please use a mouse click to enter each game. These games are independent of each other, so the results of one game do not affect the others. Before the "official" experiment begins, you can learn about the system and familiarize yourself with the logic of the game by playing the stand-alone version. When the experiment starts, you can click the game button of online version to enter. The game will be terminated when it reaches a given number of steps; at that point you will be returned to select page to select another game if it is available. While playing a game, you cannot return to the front page.

*Step 4 Play game.* Each participant will remain at this step and play one game until the game reaches the end (see Fig. S14). The interface shows his/her strategy, his/her payoff, the strategy of one of his/her neighbors, the payoff of this neighbor, the level of resource availability, and the neighbor cooperation ratio (i.e., the % of neighbors who are cooperating). Each participant can see his/her total payoffs of the whole experiment increasing step by step on the right top frame.

*Step 5 Termination of one game.* When the total number of rounds in one game (Step 4) reaches a given maximum number of time steps, the game ends. The interface will show the message "This game is finished!" and the player will be returned to game selection (Step 3). At this point the participant can select another game if it is available.

*Step 6 Conclusion.* If there is no game other on the front page, the experiment is finished.

### 3.6 Management tool

The platform manager can add, edit, reset, enable game and export game result. As an administrator, we firstly set $K = 1, T = 2$, normalized extraction parameter $\hat{e}_C = 0.7, \hat{e}_D = 1.1$, initial condition $R_0 = 0.5, x_0 = 0.5$ and game time $t_f = 40$ on the management page (see Fig. S15). Before the experiment, we disabled all games. When the appointed time is up, we enabled all games. At that point the first appeared on the front page.

## 4 Optimal control framework

Now we consider the overall payoff obtained by all players along a given time interval from $t_0$ to $t_f$, that we call cumulative payoff, that can be calculated as $\mathcal{P}(t_f) = \int_{t_0}^{t_f} TR(x\hat{e}_C + (1-x)\hat{e}_D)dt$. We note that this is proportional to the total payoff obtained on average by each player during the game.

The cumulative payoff implicitly depends on the selective parameter $w(t)$. As explained in the main text and in section 1.4, it is clear that there is a trade-off between the instantaneous payoff (maximized for $w(t) > 0$) and the cumulative payoff (that if $t_f$ is large enough it is optimized for $w(t) < 0$). In this section, we want to present a framework based on optimal control theory that allow to find the optimal $w(t)$ so to maximize the cumulative payoff given a certain potential duration of the game $t_f$.

In control theory we aim at finding the control variable and the associated state variables to maximize the given objective functional subject to dynamic constraint, initial conditions, and boundary conditions (*8-11*). In following sub-sections, we will apply such a framework to the cases that shown in section 1.4, i.e., resource dynamics with fixed strategy, strategies evolutionary dynamics with fixed resources, coupled dynamics. In particular, we aim at finding a control variable $w(t)$ and the associated state variables $R(t), x(t)$ to maximize the objective functional given by the average player cumulative payoff (we neglect the proportionality constant), i.e.,

$\max_{w(t)} \int_{t_0}^{t_f} TR(x\hat{e}_C + (1-x)\hat{e}_D)dt$ subject to dynamics constraints

$\begin{cases} \dfrac{dR}{dt} = T\left(R(1-R) - R\left(x\hat{e}_C + (1-x)\hat{e}_D\right)\right) \\ \dfrac{dx}{dt} = -wR(1-x)x \end{cases}$, initial conditions $R(0) = R_0, x(0) = x_0$ and boundary conditions $\underline{R} \le R(t) \le \overline{R}, \underline{x} \le x(t) \le \overline{x}, \underline{w} \le w(t) \le \overline{w}$ where $\_$ and $\bar{\phantom{x}}$ represents lower and upper boundary. We will solve it numerically, using the Matlab toolbox "Open Optimal Control Library" (*12*).

## 4.1 Optimal control for resource dynamics with fixed strategy

We can deal with resource dynamics with fixed strategy by setting the lower and upper boundary of variable x at its initial condition. We solve this case with growth rate $T = 2$, normalized extraction parameter $\hat{e}_C = 0.7, \hat{e}_D = 1.1$, initial condition $R_0 = 0.5, x_0 = 0.5$ and boundary condition $\underline{R} = 0, \overline{R} = 1, \underline{x} = \overline{x} = x_0, \underline{w} = -1, \overline{w} = 1$. As shown in Fig. S16, we find that $w(t)$ is always 0 to guarantee that $x(t)$ does not change and $R(t)$ always decreases and tends to reach 0. Therefore, within this framework we retrieve the results of sub-section 1.4.1.

## 4.2 Optimal control for strategies evolutionary dynamics with fixed resources

By setting the lower and upper boundary of variable R at its initial condition, we can deal with strategy dynamics with fixed resources. We solve this case with growth rate $T = 2$, normalized extraction parameter $\hat{e}_C = 0.7, \hat{e}_D = 1.1$, initial condition $R_0 = 0.5, x_0 = 0.5$ and boundary condition $\underline{R} = \overline{R} = R_0, \underline{x} = 0, \overline{x} = 1, \underline{w} = -1, \overline{w} = 1$. As shown in Fig. S17 we find that, in the absence of environmental feedback the most rewarding strategy is always attained with defection, i.e., $w(t) > 0$. In fact, in the above condition $x$ always decrease and tend to reach 0 as soon as possible to maximize $x\hat{e}_C + (1-x)\hat{e}_D$ at each time step. Independently of $t_f$ the optimal strategy always leads to depletion of resources. Therefore, within this framework we retrieve the results of sub-section 1.4.2.

## 4.3 Optimal control for coupled dynamics

We now consider the full HES dynamics with environmental feedback. We aim at finding a control variable $w(t)$ and the associated state variables $R(t), x(t)$ so to maximize the objective functional $\max_{w(t)} \int_{t_0}^{t_f} TR(x\hat{e}_C + (1-x)\hat{e}_D)dt$ subject to dynamics constraints

$$\begin{cases} \dfrac{dR}{dt} = T\left(R(1-R) - R\left(x\hat{e}_C + (1-x)\hat{e}_D\right)\right) \\ \dfrac{dx}{dt} = -wR(1-x)x \end{cases}$$ , initial conditions $R(0) = R_0, x(0) = x_0$ and boundary

conditions $0 \le R(t) \le 1, 0 \le x(t) \le 1, -1 \le w(t) \le 1$.

We now summarize the main results of the optimal control framework depending on players expectation on game duration. If players assume that $t_f - t_0$ is small, then the time horizon of their maximization problem is short, and they repeated the optimal control framework piecewise for each time step, i.e., by maximizing the objective functional $\max_{w(t)} \int_{i}^{i+1} TR(t)\left(x(t)\hat{e}_C + \left(1-x(t)\right)\hat{e}_D\right)dt$ for $i = 0,\ldots,t_f - 1$. In particular, in the first time step players maximize the objective functional $\max_{w(t)} \int_0^1 TR(x\hat{e}_C + (1-x)\hat{e}_D)dt$ subject to the dynamics constraints, initial conditions $R(0) = R_0, x(0) = x_0$ and boundary conditions. In the second time step players will maximize the objective functional $\max_{w(t)} \int_1^2 TR(x\hat{e}_C + (1-x)\hat{e}_D)dt$ subject to the dynamic's constraints, initial conditions $R(1) = R_0, x(1) = x_0$ and boundary conditions. In this case we find that resource $R(t)$ is soon depleted and the *Tragedy of the Commons* is retrieved (see Fig. S18 and S19). In other words, this strategy is equivalent to the case of maximization of instantaneous payoff (fixed $w > 0$), if players behave disregarding any future projections of their actions in the resources. Conversely, if players assume that $t_f - t_0$ is not small, then the time horizon of their maximization problem is not short, and they will consider optimal control framework continuously for a relative long time. For intermediate values of $t_f - t_0$ (see Fig. S20 and S21), the optimal solution first gives $w(t) = -1$ (all players quickly adopt cooperation), then shows a sudden shift to $w(t) = 1$ (all players switch to defection) thereby leading to a corresponding relevant decrease of resources, yet avoiding CPR collapse. Finally, if $t_f - t_0$ is relatively large (see Fig. S20 and S21), then the solutions of coupled dynamics imply that players self-organize adopting long term cooperation. Once complete cooperation is reached ($x(t) = 1$), players do not change strategy ($w(t) = 0$) until just before the game ends, when all players abruptly move to defection ($w(t) = 1$). This change, however, happens only just before the end of the game, and it has no actual effect on resources sustainability.

# 5 Inferring players optimal strategies from the experimental data

We can define the error between trajectory of experiment and optimal control as an indicator of performance of the players with respect to the optimal strategy, i.e., $\mathcal{E} = \sum_{t=0}^{t_f} \sqrt{\left(x_{exp}(t) - x_{opt}(t)\right)^2}$ . Let's now define the critical time $\tau_{crt}$ as the time needed by the players to self-organize and reach a desired state. Specifically, for $0 \leq t \leq \tau_{crt}$ players consider in predicting their behavior so to achieve their desired state, so self-organization emerges and we use optimal control for coupled dynamics (piecewise version for game type 1 and continuous version for game type 2) to infer $w(t)$ ; for $\tau_{crt} \leq t \leq t_f$ players reach a steady state self-organized cooperation level and we implement the optimal control constrained by the resource dynamics, but with fixed strategy. To infer the value of $\tau_{crt}$ we thus first calculate the indicator $\mathcal{E}$ for $\tau = 1, \ldots, t_f$ and then see which of such values minimizes the error indicator $\mathcal{E}$ . In Fig. S22 we show the error indicators as a function of different time horizons $\tau$ for the experimental data we collected. We find that the critical time $\tau_{crt}$ of game type 1 are 17, 14, 8, for experiments carried out in China, and 1, 1, 4 for those carried out in USA. For game type 2 the critical time $\tau_{crt}$ are 9, 11, 9, and 4, 3, 5, for experiments in China and in USA, respectively.

# 6 Experimental results

As explained in the main text and in the above sections, we run three experiments in China with players interactions given by a complete network, a BA network and a SW network with size $N = 20$ and connectivity $C = 0.2$ and three experiments in the USA where players interact via complete networks. Fig. S23 shows related results for game of type 1 and Fig. S24 shows the results for type 2 games. Additionally, we also compare the average individual cumulative payoff of between experimental result and theoretical prediction (Fig. S25). The results show an extremely good fit of the experimental data by our model.

# References and Notes

1. M. A. Nowak, A. Sasaki, C. Taylor, D. Fudenberg, Emergence of cooperation and evolutionary stability in finite populations. *Nature* **428**, 646-650 (2004).
2. M. A. Nowak, *Evolutionary dynamics: exploring the equations of life*. (Harvard university press, 2006).
3. A. R. Tilman, J. B. Plotkin, E. Akçay, Evolutionary games with environmental feedbacks. *Nature communications* **11**, 1-11 (2020).
4. E. A. Coayla-Teran, S.-E. A. Mohammed, P. R. C. Ruffino, Hartman-Grobman theorems along hyperbolic stationary trajectories. *Discrete & Continuous Dynamical Systems* **17**, 281 (2007).
5. O. Filipova, *Learning Vue. js 2*. (Packt Publishing Ltd, 2016).
6. S. Tilkov, S. Vinoski, Node. js: Using JavaScript to build high-performance network programs. *IEEE Internet Computing* **14**, 80-83 (2010).
7. D. Crockford, The application/json Media Type for JavaScript Object Notation (JSON),(2006). *URL http://tools.ietf.org/html/rfc4627* **38**, (2006).
8. E. Kreyszig, *Introductory functional analysis with applications*. (wiley New York, 1978), vol. 1.
9. S. Lenhart, J. T. Workman, *Optimal control applied to biological models*. (Chapman and Hall/CRC, 2007).
10. F. L. Lewis, D. Vrabie, V. L. Syrmos, *Optimal control*. (John Wiley & Sons, 2012).
11. J. B. Conway, *A course in functional analysis*. (Springer, 2019), vol. 96.
12. J. Koenemann, G. Licitra, M. Alp, M. Diehl, *Openocl–open optimal control library*. (2017).

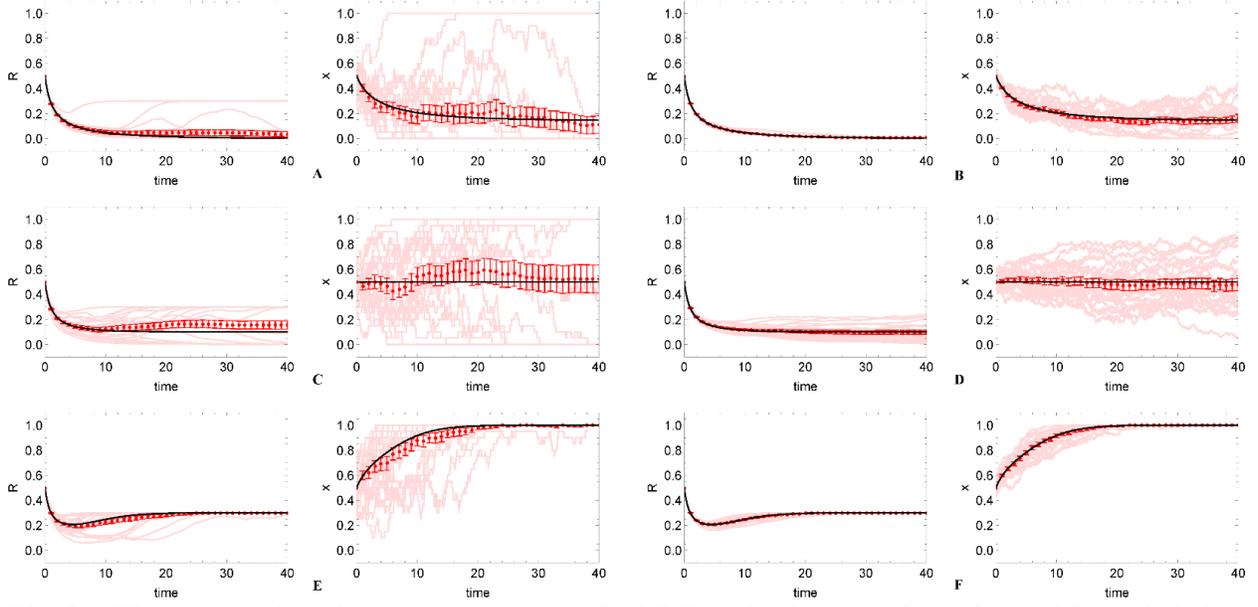

**Fig. S1. The comparison between macroscopic ODE and microscopic update with varied size.**
(**A**) $N = 20, w = 1$ ; (**B**) $N = 200, w = 1$ ; (**C**) $N = 20, w = 0$ ; (**D**) $N = 200, w = 0$ ; (**E**) $N = 20, w = -1$ ; (**F**) $N = 200, w = -1$ . Parameters in these games are growth rate $T = 2$ , normalized extraction parameter $\hat{e}_C = 0.7, \hat{e}_D = 1.1$ and initial condition $R_0 = 0.5, x_0 = 0.5$ . Each light red line is one realization of numerical simulation and the black line is the theoretical prediction of macroscopic ODE. The red points and red fences are the average and stand error of all realizations at each integer time.

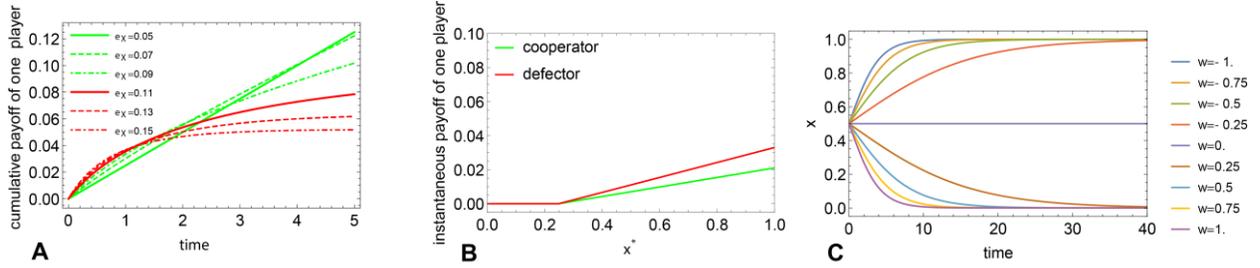

**Fig. S2**. **Behavior of single resource dynamics and single evolutionary dynamics.** (**A**) The average cumulative payoff earned by one player at the end of the game for different extraction rates $e_X$ where $T = 2, N = 20$ and $R_0 = 0.5$. The green curves represent cooperation (i.e., strategy $X = C$), while the red curves represent defection (i.e., strategy $X = D$); (**B**) Comparison between instantaneous payoff of one cooperator and defector after the system reaches stable equilibrium for the case with a constant strategy $x = x^*$ and with $0 < x^* < 1$ (here we used $\hat{e}_C = 0.7, \hat{e}_D = 1.1$, and thus $x_l = 0.25$). (**C**) Trajectory of single evolutionary dynamics for different values of $w$ where $T = 2$, $\hat{e}_C = 0.7$, $\hat{e}_D = 1.1$, $R_0 = 0.5$ and $x_0 = 0.5$.

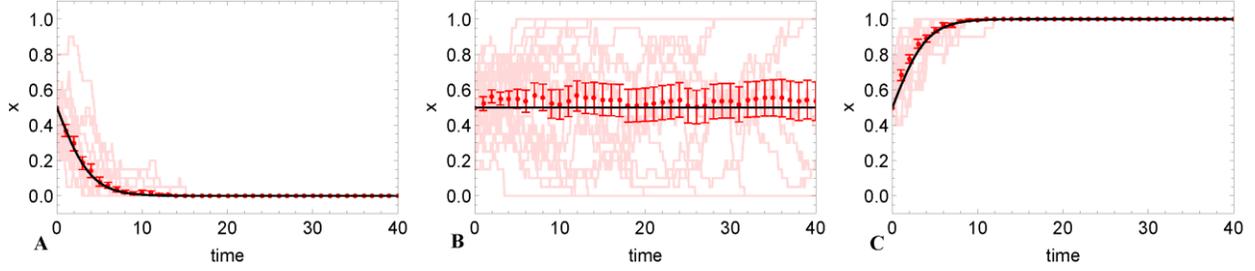

**Fig. S3. The comparison between macroscopic ODE and microscopic dynamics with fixed resources.** (**A**) $w = 1$ ; (**B**) $w = 0$ ; (**C**) $w = -1$. Parameters in these games are size $N = 20$, growth rate $T = 2$, normalized extraction parameter $\hat{e}_C = 0.7, \hat{e}_D = 1.1$ and initial condition $R_0 = 0.5, x_0 = 0.5$. Each light red line is one realization of numerical simulation and black line is the theoretical prediction of macroscopic ODE. The red points and red fences are the average and stand error of all realizations at each integer time.

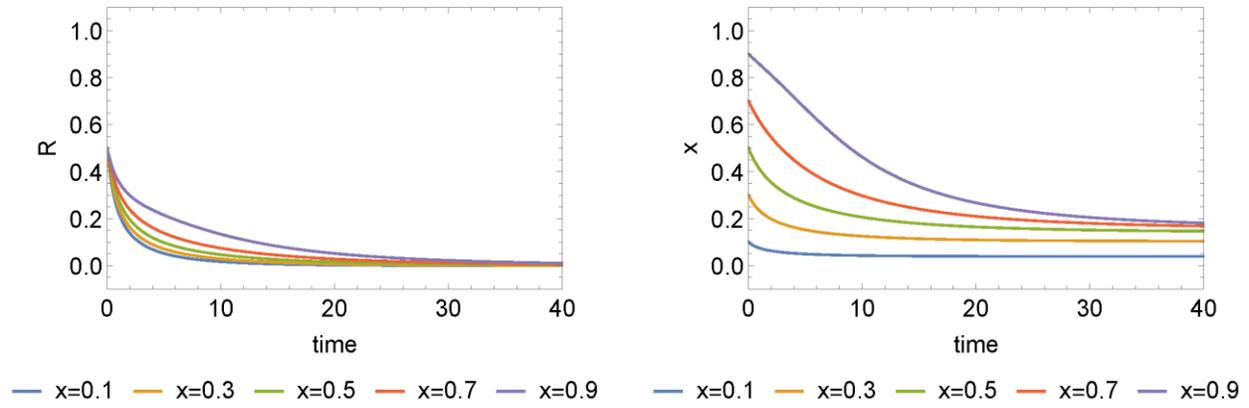

**Fig. S4**. **The trajectory of $R$ and $x$ with different initial condition** $x(0) = 0.1, 0.3, 0.5, 0.7, 0.9$ **and fixed** $R(0) = 0.5$. Parameters in these games are growth rate $T = 2$, normalized extraction parameter $\hat{e}_C = 0.7, \hat{e}_D = 1.1$.

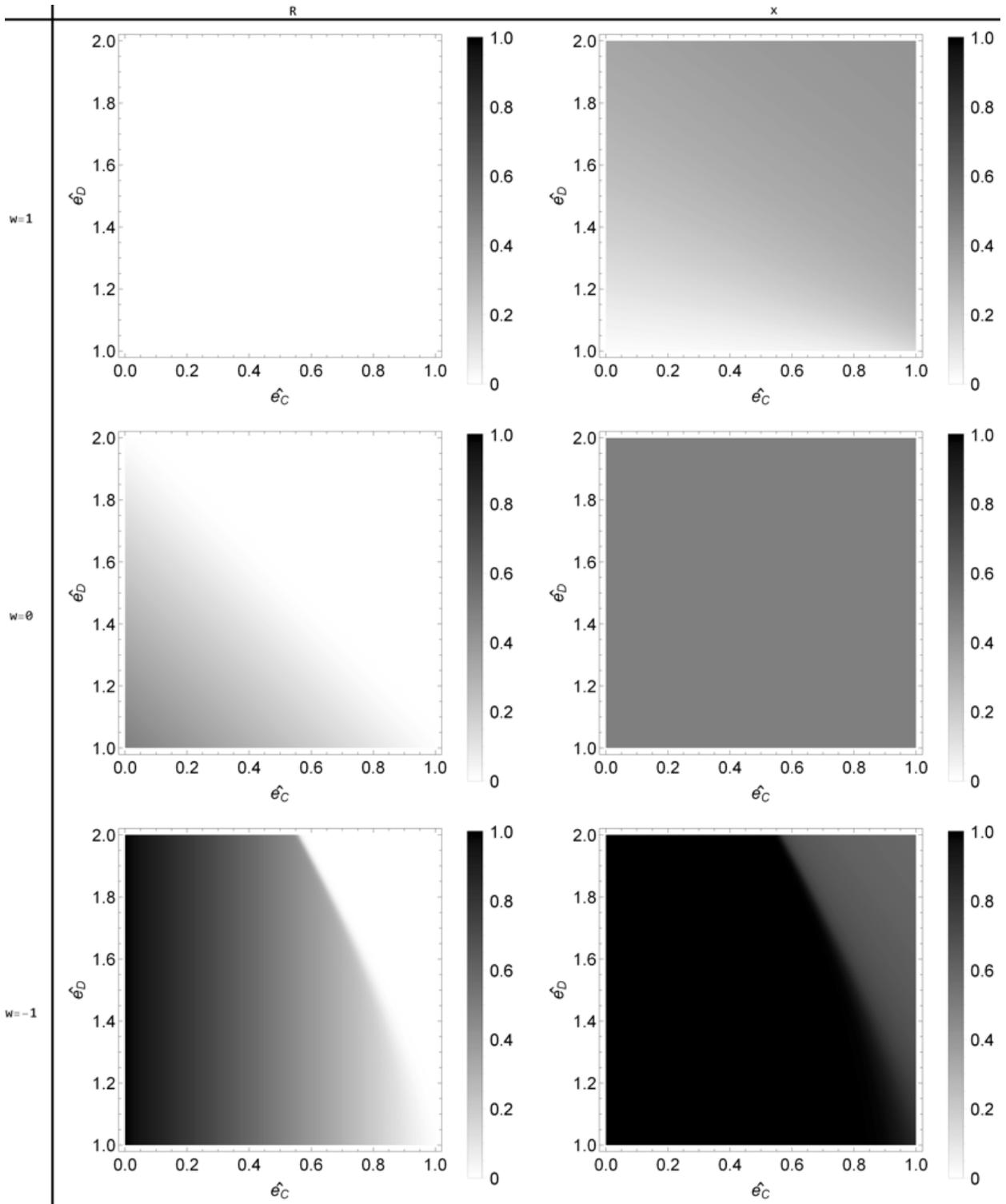

**Fig S5**. **The stable equilibrium of** $R$ **and** $x$ **with** $w = 1, 0, -1$. Parameters in these games are growth rate $T = 2$, normalized extraction parameter $\hat{e}_C = 0.7, \hat{e}_D = 1.1$ and initial condition $R_0 = 0.5, x_0 = 0.5$.

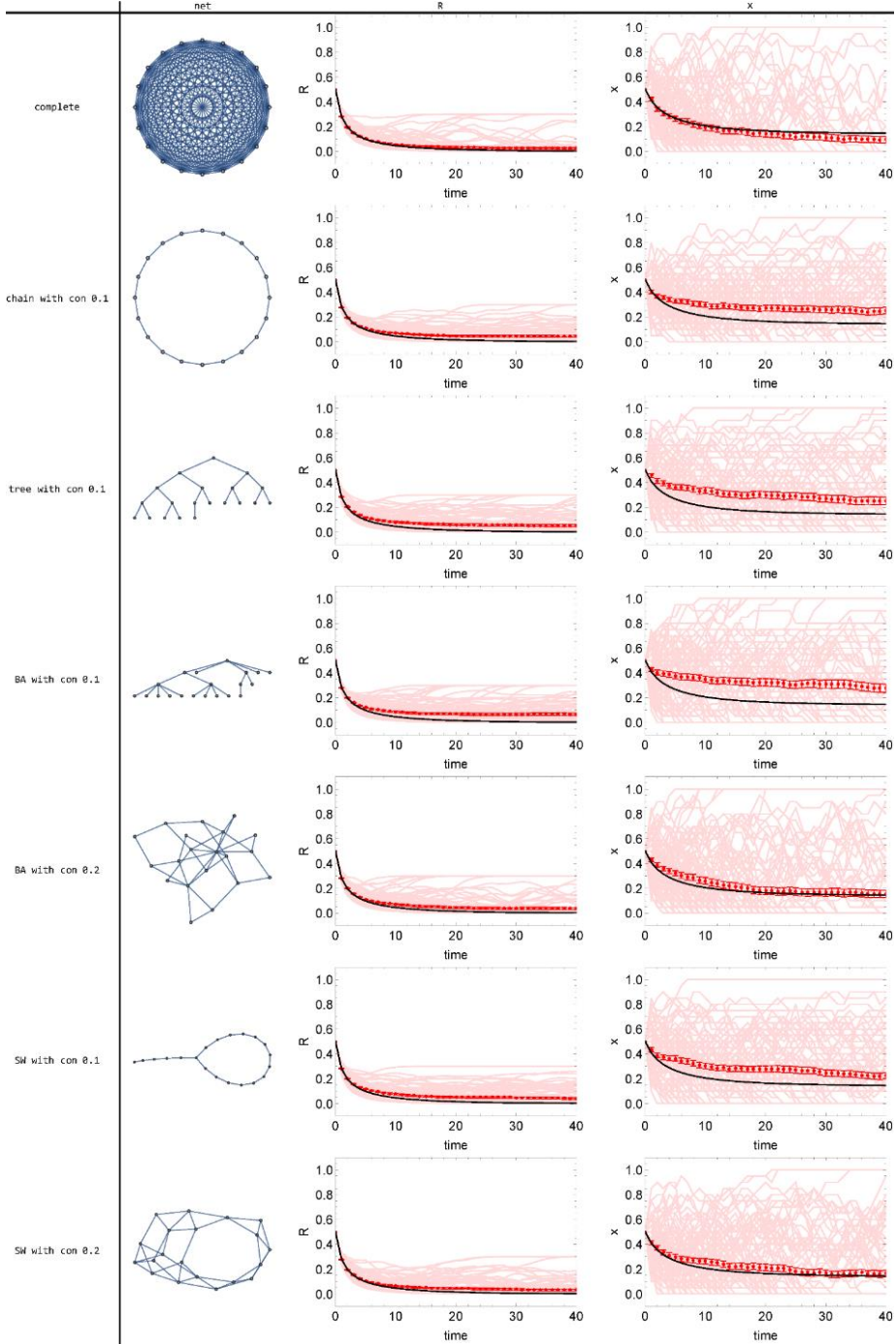

**Fig. S6**. **HES dynamics with** $w = 1$ **with different social network structures.** From the top: complete network, chain network with connectivity C=0.1, tree network with C=0.1, Barabasi-Albert (BA) network with C=0.1, BA network with C=0.2, Small-World (SW) network with C=0.1, SW network with C=0.2. Parameters in these games are size $N = 20$, growth rate $T = 2$, selection pressure parameter $w = 1$, game time $t_f = 40$, normalized extraction parameter $\hat{e}_C = 0.7, \hat{e}_D = 1.1$ and initial condition $R_0 = 0.5, x_0 = 0.5$. Each light red line is one realization of numerical simulation and black line is the theoretical prediction of macroscopic ODE. The red points and red fences are the average and stand error of all realizations at each integer time.

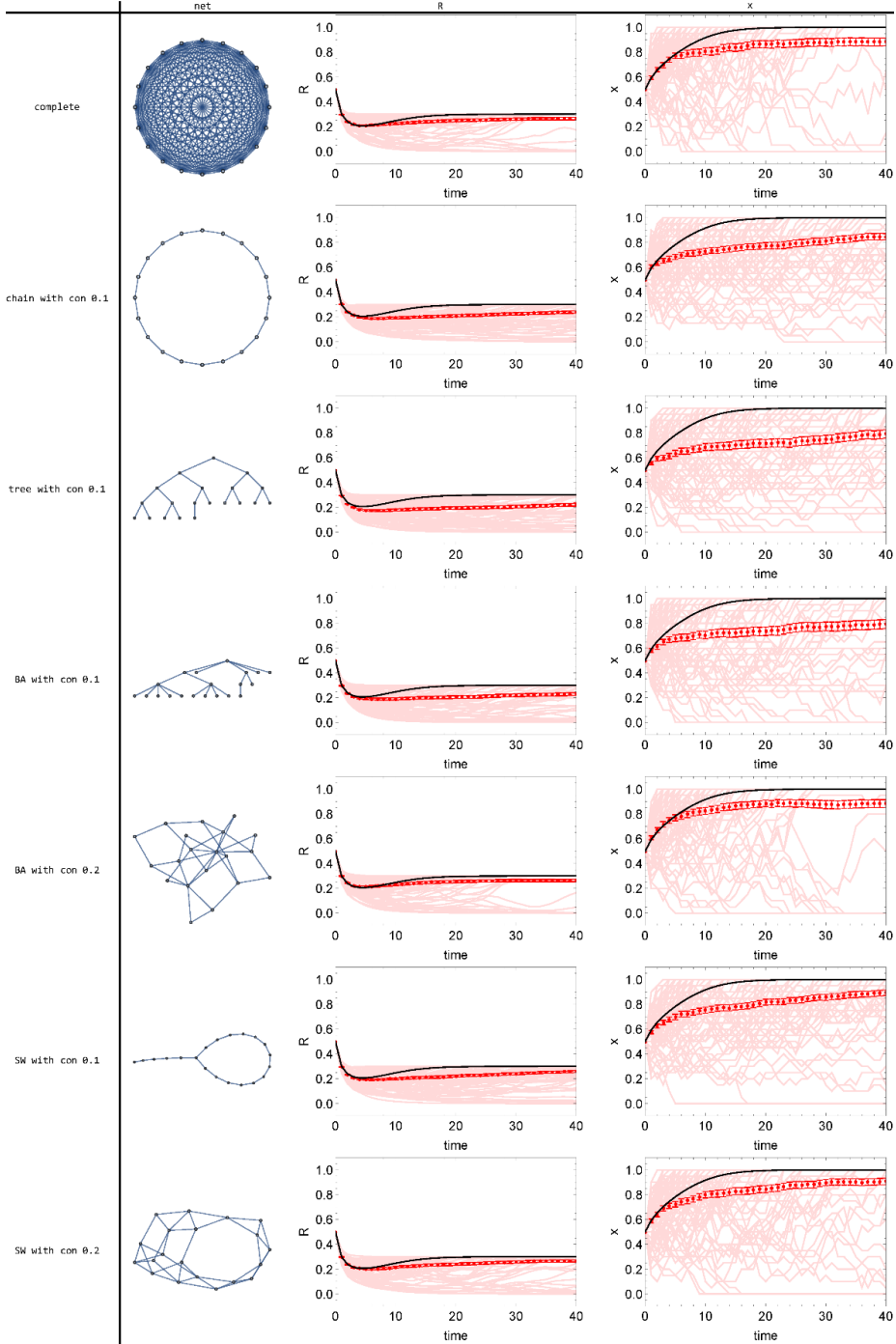

**Fig. S7**. **HES dynamics with $w = -1$ with different social network structures.** From the top: complete network, chain network with connectivity C=0.1, tree network with C=0.1, Barabasi-Albert (BA) network with C=0.1, BA network with C=0.2, Small-World (SW) network with C=0.1, SW network with C=0.2. Parameters in these games are size $N = 20$, growth rate $T = 2$, selection pressure parameter $w = -1$, game time $t_f = 40$, normalized extraction parameter $\hat{e}_C = 0.7, \hat{e}_D = 1.1$ and initial condition $R_0 = 0.5, x_0 = 0.5$. Each light red line is one realization of numerical simulation and black line is the theoretical prediction of macroscopic ODE. The red points and red fences are the average and stand error of all realizations at each integer time.

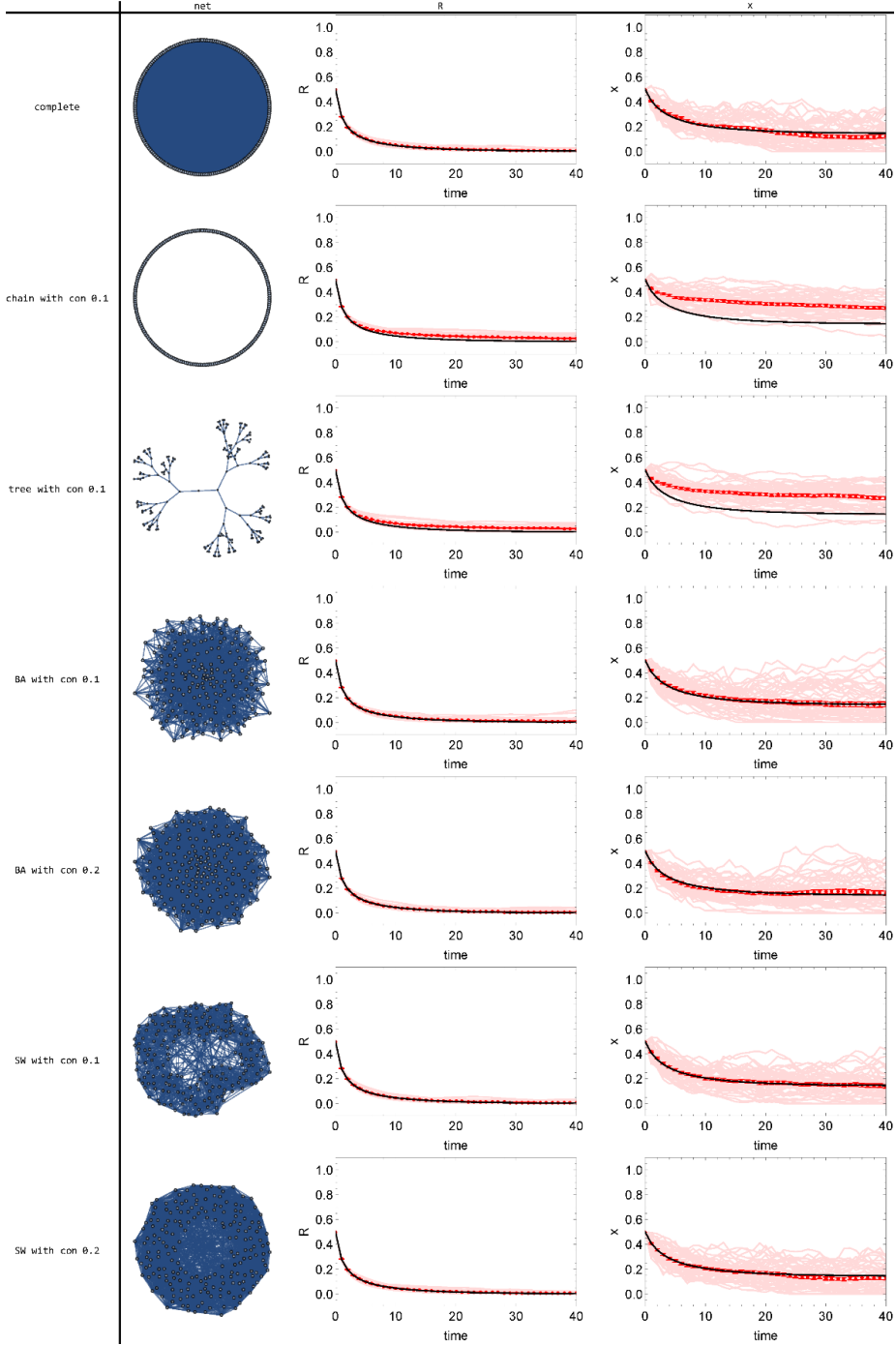

**Fig. S8**. **HES dynamics with** $w = 1$ **with different social network structures.** From the top: complete network, chain network with connectivity C=0.1, tree network with C=0.1, Barabasi-Albert (BA) network with C=0.1, BA network with C=0.2, Small-World (SW) network with C=0.1, SW network with C=0.2. Parameters in these games are size $N = 200$, growth rate $T = 2$, selection pressure parameter $w = 1$, game time $t_f = 40$, normalized extraction parameter $\hat{e}_C = 0.7, \hat{e}_D = 1.1$ and initial condition $R_0 = 0.5, x_0 = 0.5$. Each light red line is one realization of numerical simulation and black line is the theoretical prediction of macroscopic ODE. The red points and red fences are the average and stand error of all realizations at each integer time.

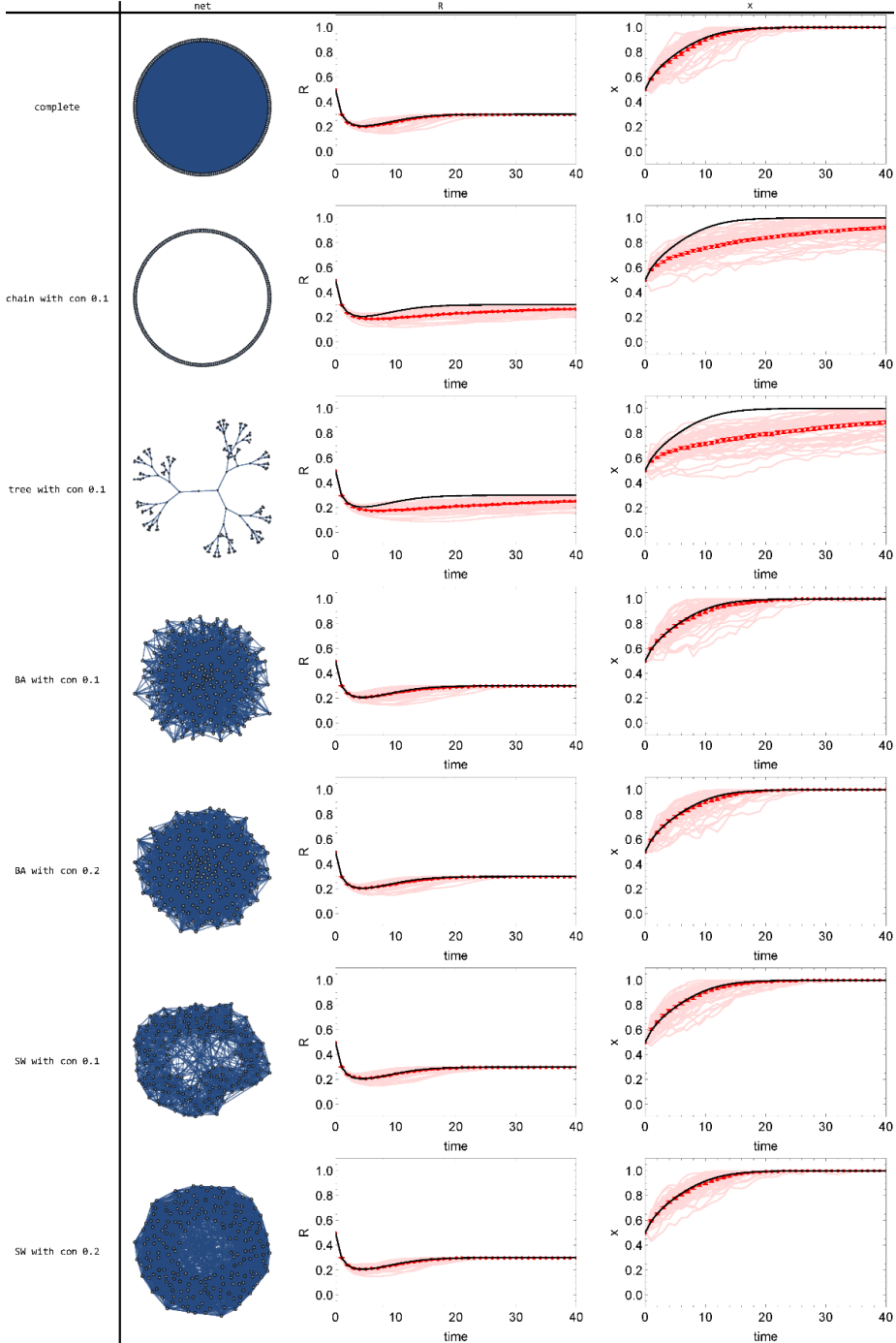

**Fig. S9**. **HES dynamics with** $w = -1$ **with different social network structures.** From the top: complete network, chain network with connectivity C=0.1, tree network with C=0.1, Barabasi-Albert (BA) network with C=0.1, BA network with C=0.2, Small-World (SW) network with C=0.1, SW network with C=0.2. Parameters in these games are size $N = 200$, growth rate $T = 2$, selection pressure parameter $w = -1$, game time $t_f = 40$, normalized extraction parameter $\hat{e}_C = 0.7, \hat{e}_D = 1.1$ and initial condition $R_0 = 0.5, x_0 = 0.5$. Each light red line is one realization of numerical simulation and black line is the theoretical prediction of macroscopic ODE. The red points and red fences are the average and stand error of all realizations at each integer time.

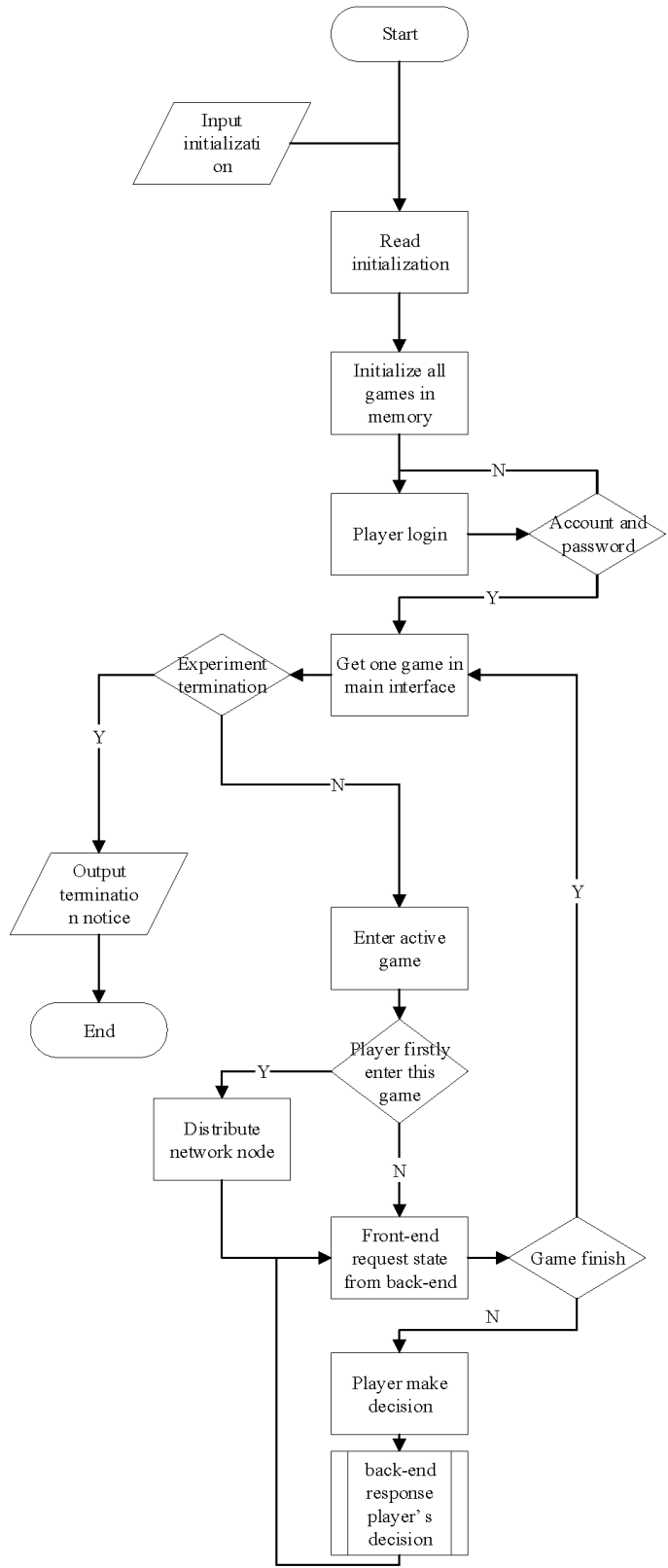

**Fig. S10. Software flow chart of interactive platform.**

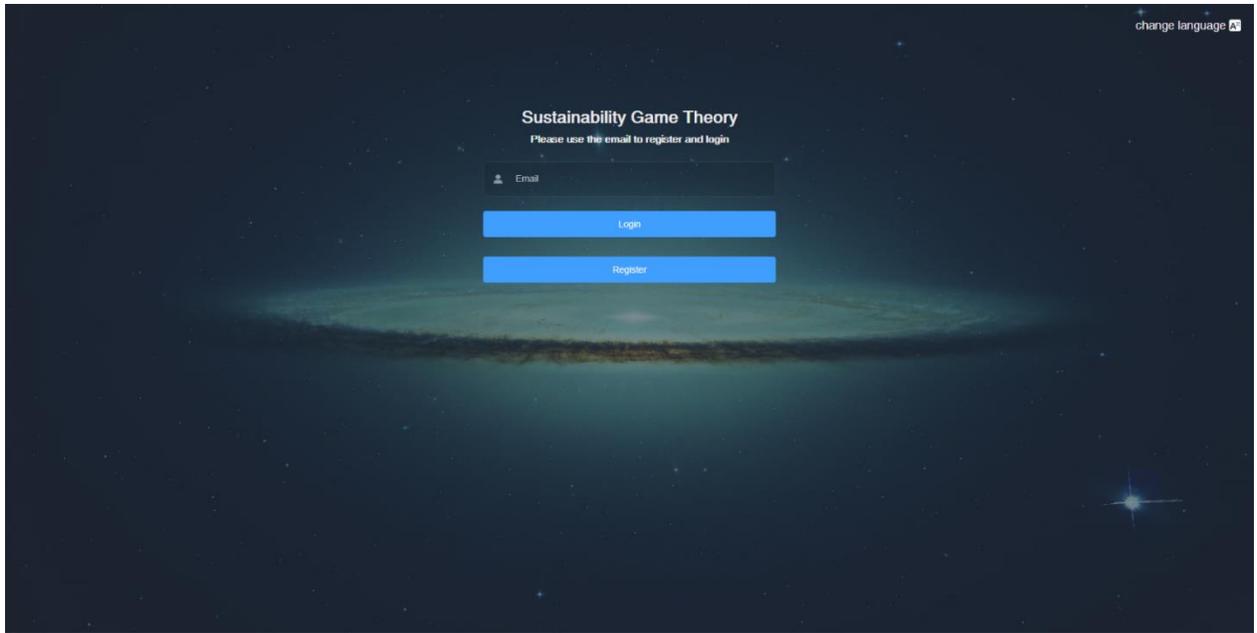
**Fig. S11. Register/Login page of the interactive platform.**

**Instructions**

The following summarizes experimental process of game type 1 and 2 step by step.

**Step 1 Register/Login**

If you have registered the interactive platform, you should login and jump to step 2; if not, you should register an account (see Fig. 1).

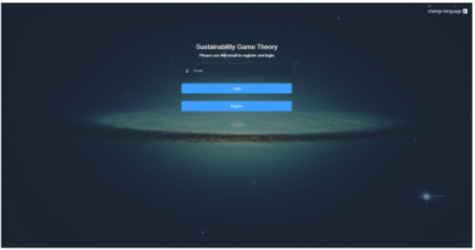

Figure 1. Register/Login page of interactive platform.

**Step 2 Understand game**

You are now participating to an interesting experiment that includes many different games. Please read the following instructions carefully. A good understanding of these instructions will allow you to perform better in the experiment and receive a better payoff.

You will play a game in which you act as a member of a community that relies on a common-pool resource (for instance a group of herders using the same pastures or a fishing community relying on the same fisheries). Every player can harvest/extract resources from the common pool. Each player needs to decide on the amount of resource he/she wants to extract by choosing between two options (or "strategies"), namely, "cooperation" and "defection". The payoff of each player will be proportional to the amount of resources that player has extracted during the game. Each game will entail several time steps and at every round you will have to choose between these two possible strategies (i.e., cooperation and defection). If you select defection, you extract a greater amount of resources than what it is (on average) sustainable. This means that if the other players do the same, then the resource will be exhausted/depleted. Conversely, if you select cooperation, you extract a smaller amount of resource. In this case the resource will not be exhausted, provided that the other players do the same (i.e., cooperate). If in the course of the

**Fig. S12. Instruction page of the interactive platform.**

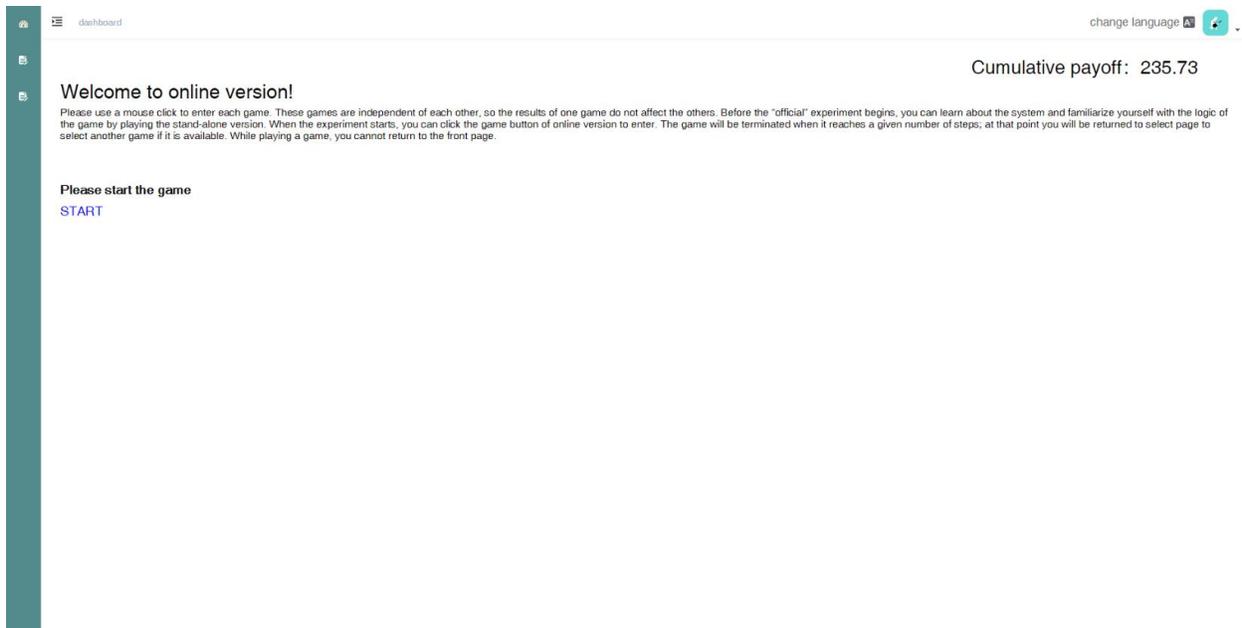

**Fig. S13. Front page of the interactive platform.**

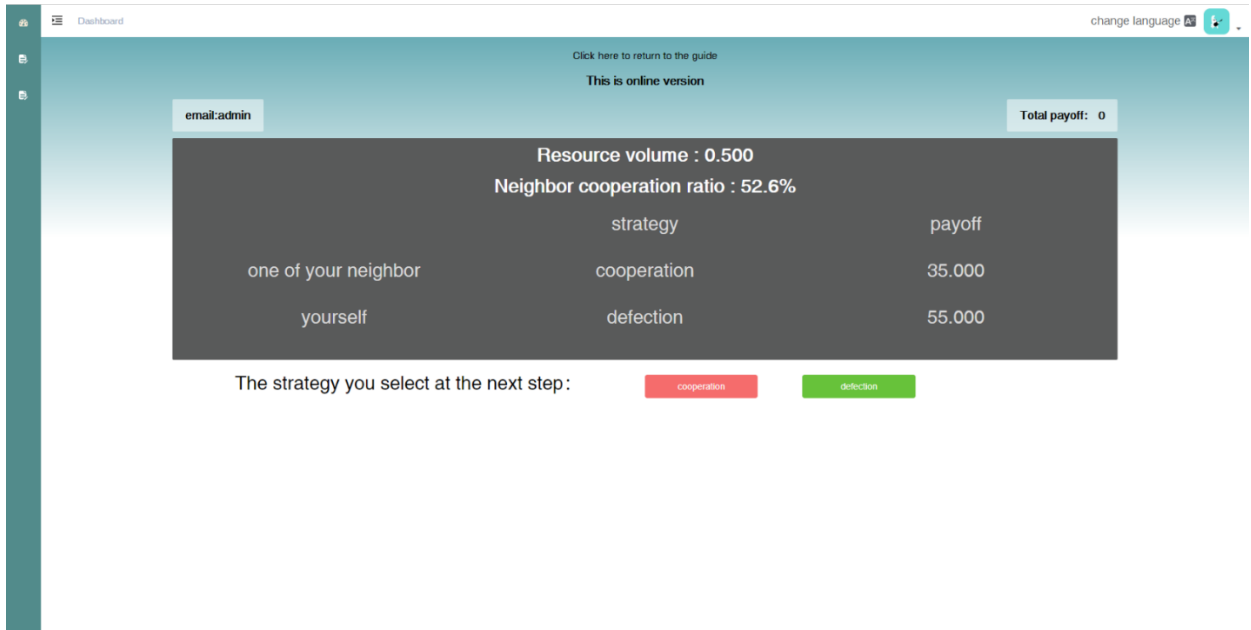

**Fig. S14. Play page of the interactive platform.**

**Fig. S15. Management page of the interactive platform.**

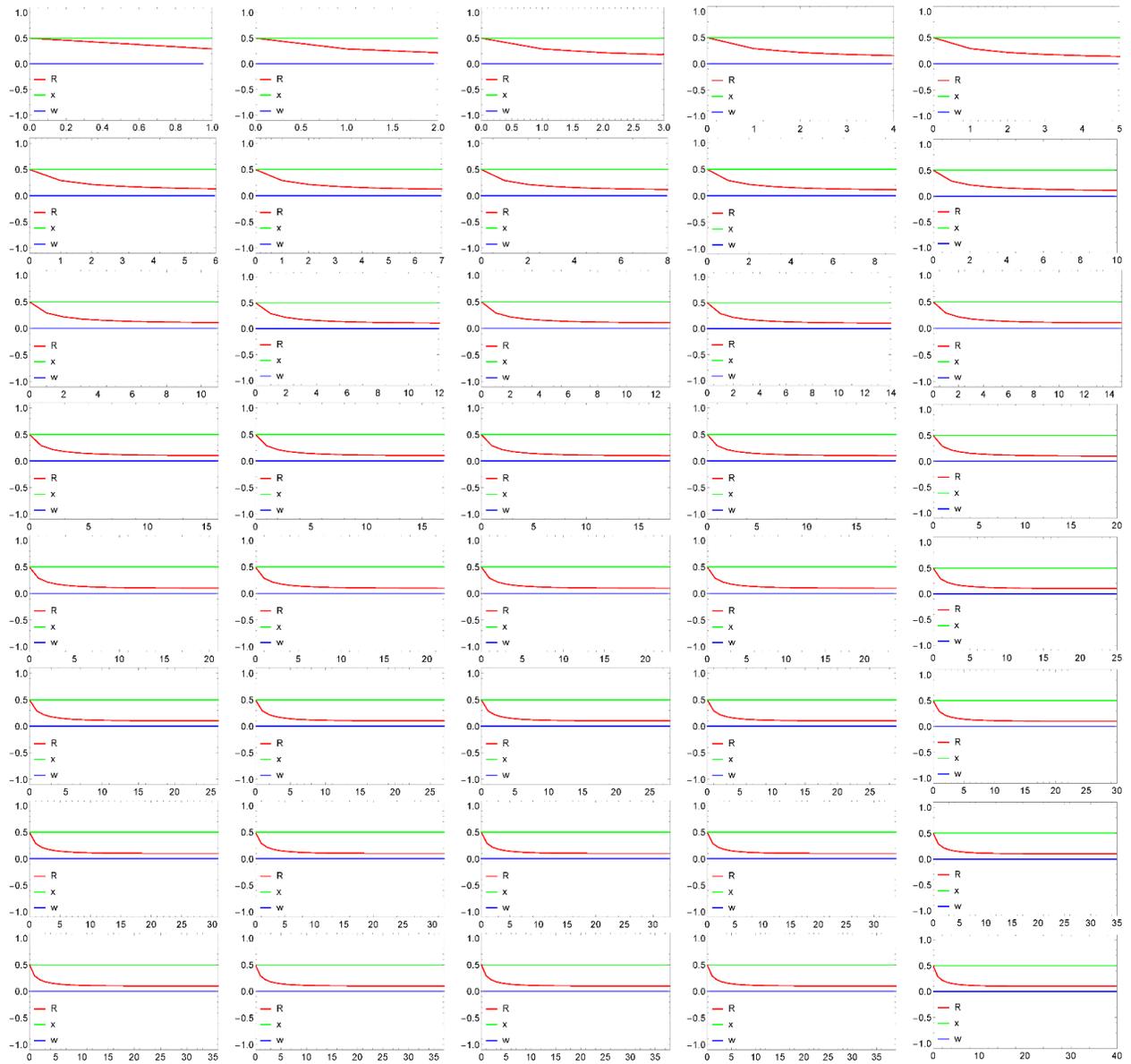

**Fig. S16. The comparison of trajectories obtained using different values of** $t_f$ **(from 1 to 40)**
**and optimal control for resource dynamics with fixed strategy.**

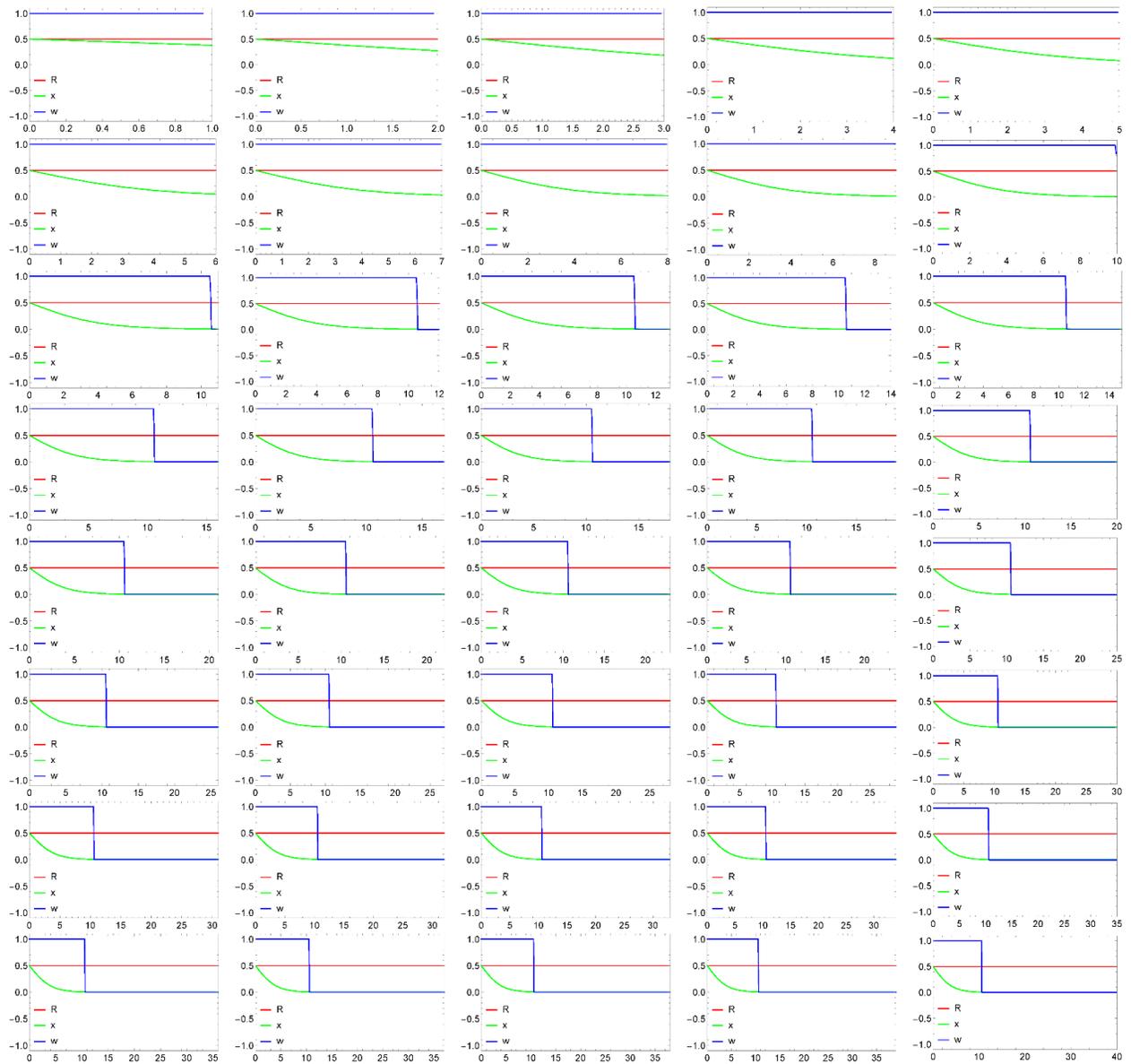

**Fig. S17.** The comparison of trajectories obtained using different values of $t_f$ (from 1 to 40) and optimal control for strategies evolutionary dynamics with fixed resources.

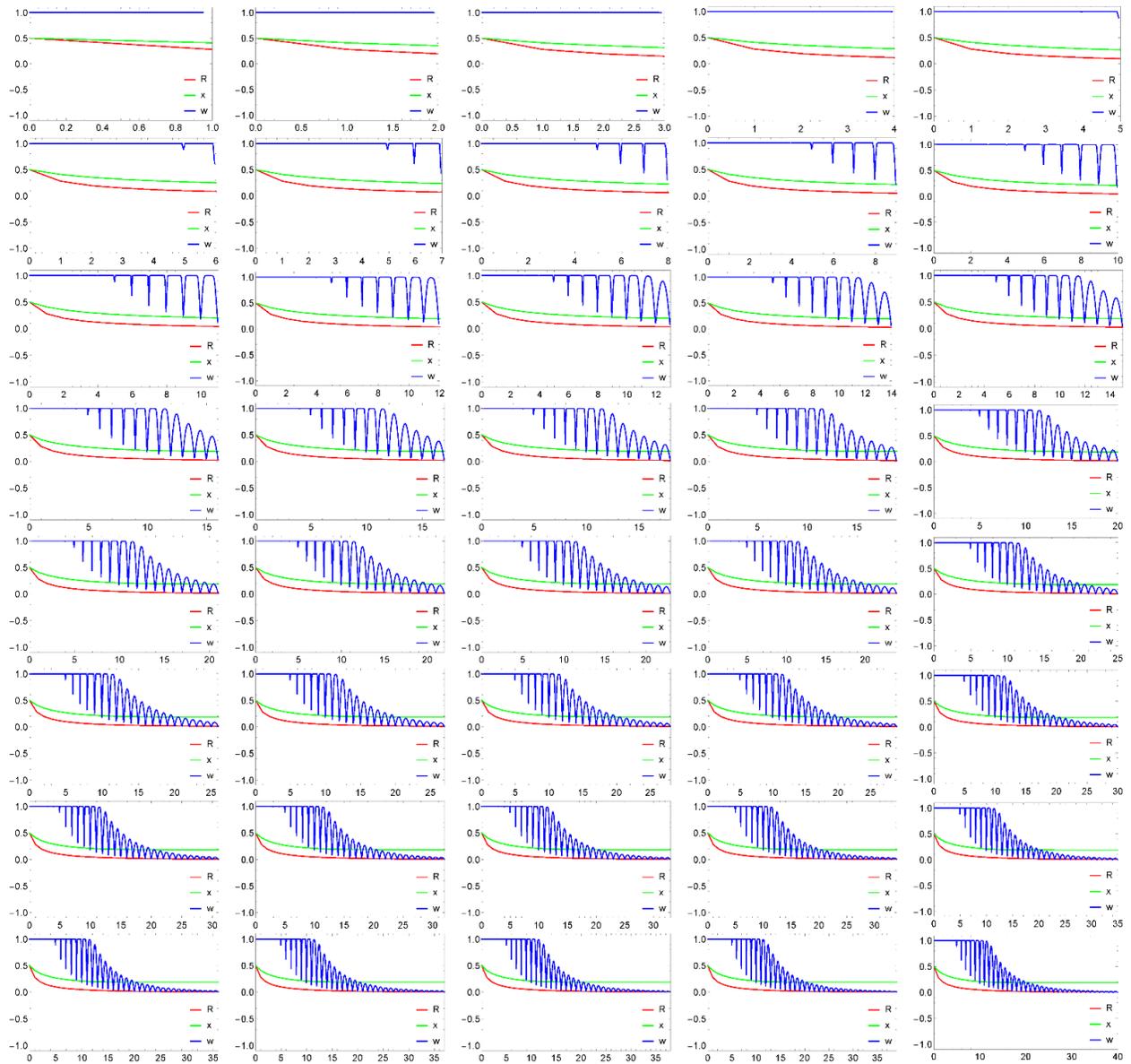

**Fig. S18. The comparison of trajectories obtained using different values of** $t_f$ **(from 1 to 40) and the piecewise optimal control for the coupled dynamics.**

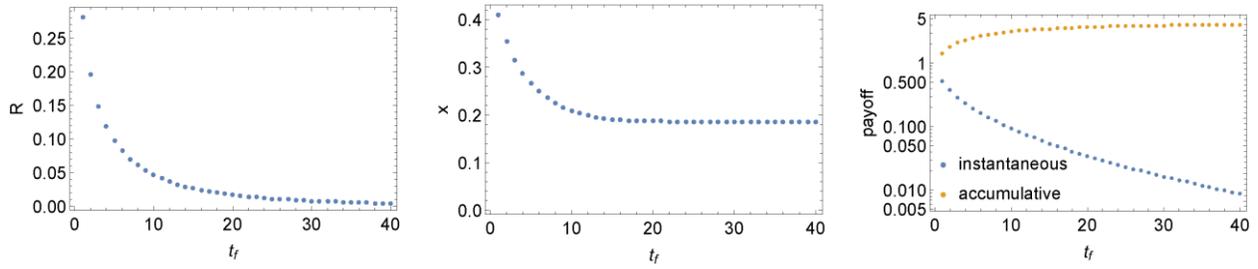

**Fig. S19. The final value of $R$, $x$, instantaneous and accumulative payoff as a function of $t_f$ using the piecewise optimal control for the coupled dynamics.**

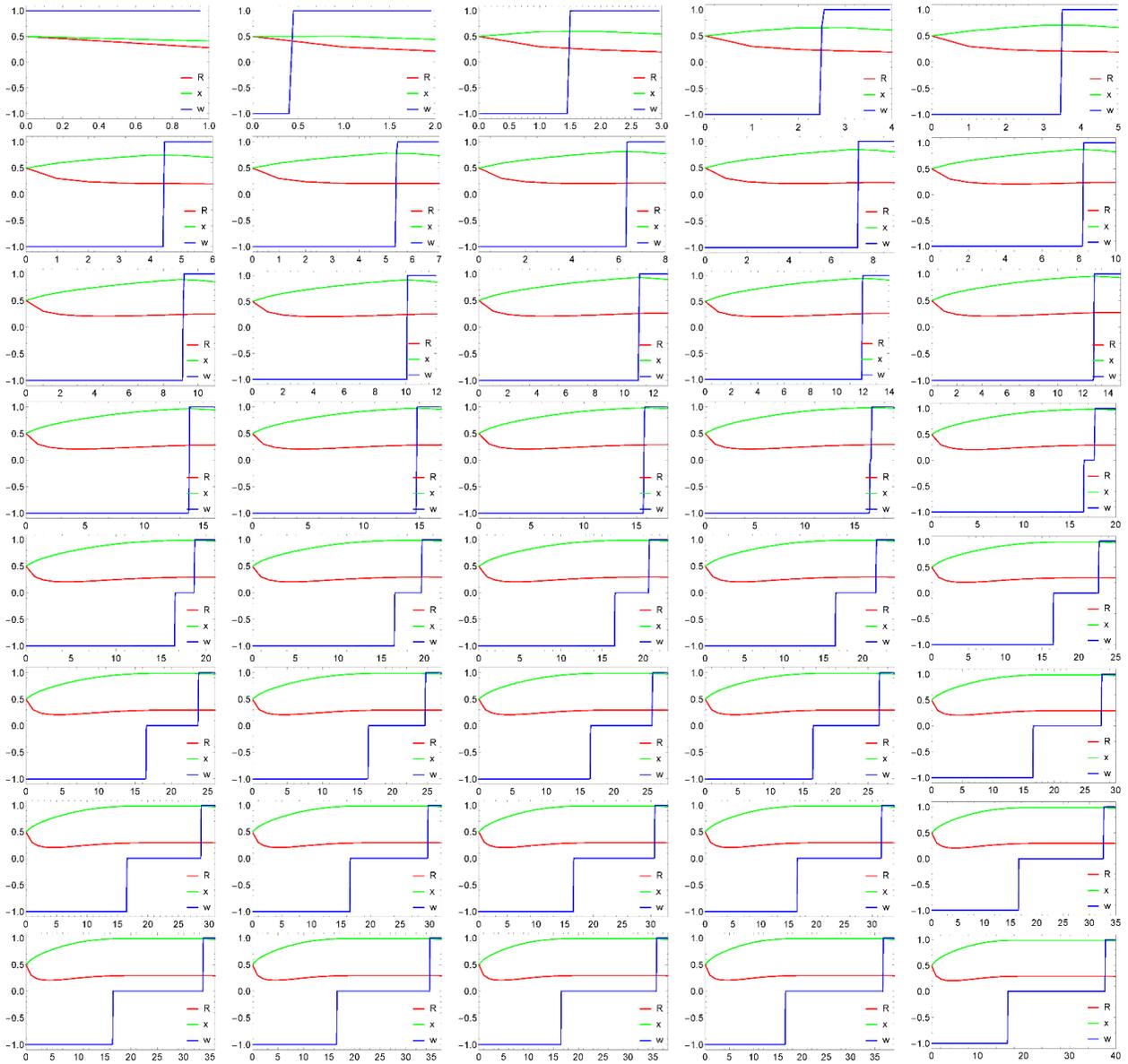

**Fig. S20. The comparison of trajectories obtained using different values of $t_f$ (from 1 to 40) and the continuous optimal control for the coupled dynamics.**

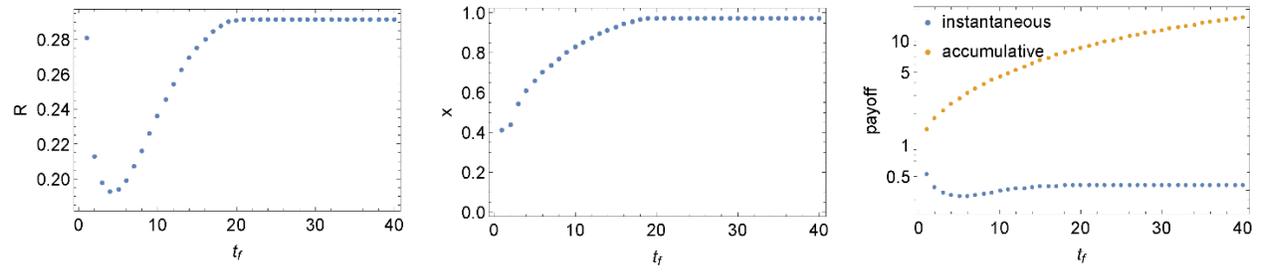

**Fig. S21. The final value of** $R$ **,** $x$ **, instantaneous and accumulative payoff as a function of** $t_f$ **using the continuous optimal control for the coupled dynamics.**

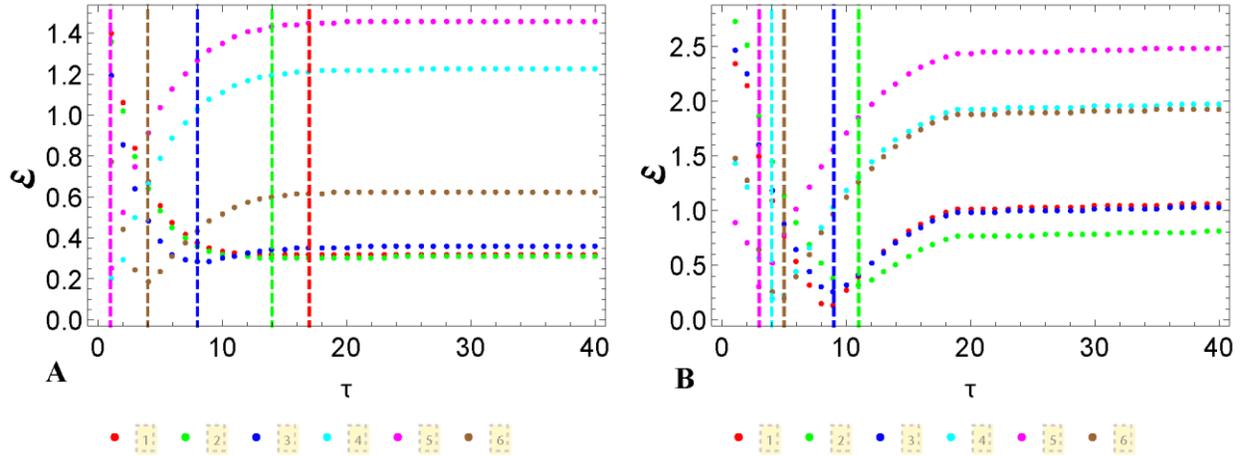

**Fig. S22. Values of error indicator** $\mathcal{E}$ **as a function of time horizon** $\tau$ **for optimal control for coupled dynamics.** (**A**) game type 1; (**B**) game type 2. The points are error indicator $\mathcal{E}$ with varied time $\tau$ and dashed lines represent critical time $\tau_{crt}$. Different colors represent different experiments including (1) complete network run in China, (2) BA network run in China, (3) SW network run in China, (4) complete network run in USA, (5) complete network run in USA, (6) complete network run in USA.

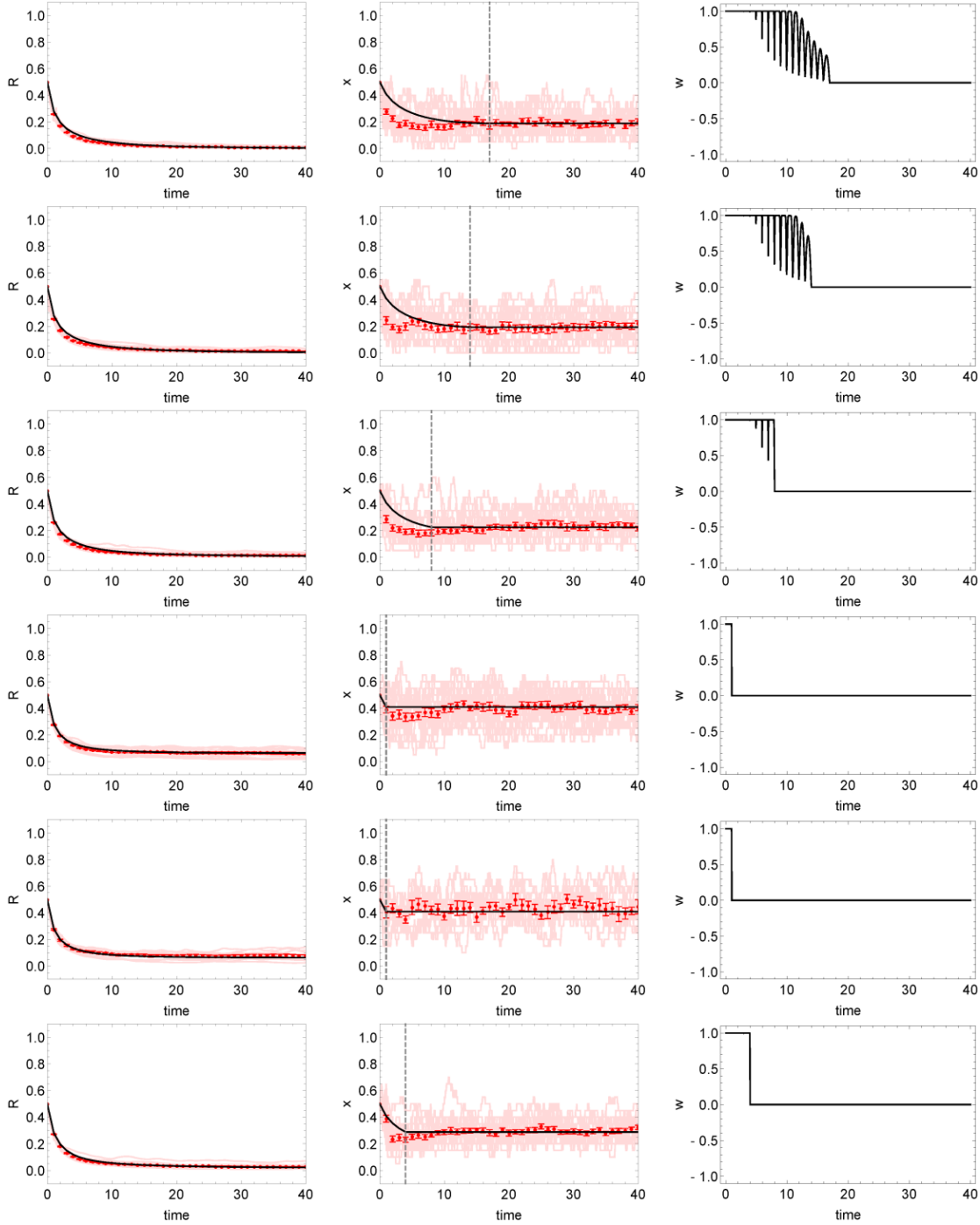

**Fig. S23. The comparison between experimental result and theoretical prediction for game type 1.** Parameters in these games are size $N = 20$, growth rate $T = 2$, normalized extraction parameter $\hat{e}_C = 0.7, \hat{e}_D = 1.1$, initial condition $R_0 = 0.5, x_0 = 0.5$ and game time $t_f = 40$. From the top: complete network run in China; BA network run in China; SW network run in China; complete network run in USA; complete network run in USA; complete network run in USA. Light red lines represent single realizations of the experiment. The red points and bars are the corresponding averages and standard errors at each integer time. The black lines represent theoretical predication. The dashed lines represent critical time $\tau_{crt}$.

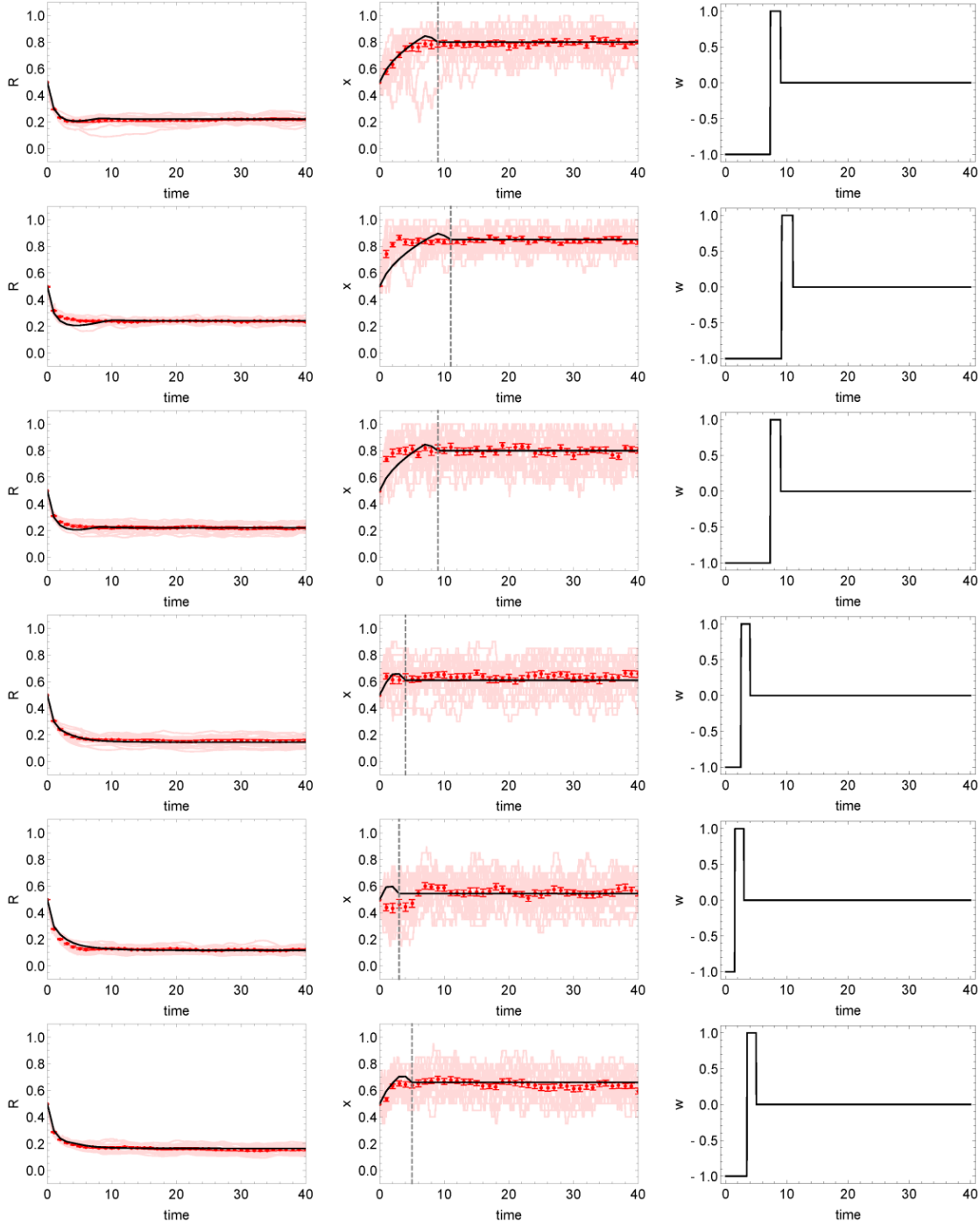

**Fig. S24.** **The comparison between experimental result and theoretical prediction for game type 2.** Parameters in these games are size $N = 20$, growth rate $T = 2$, normalized extraction parameter $\hat{e}_C = 0.7, \hat{e}_D = 1.1$, initial condition $R_0 = 0.5, x_0 = 0.5$ and game time $t_f = 40$. From the top: complete network run in China; BA network run in China; SW network run in China; complete network run in USA; complete network run in USA; complete network run in USA. Light red lines represent single realizations of the experiment. The red points and bars are the corresponding averages and standard errors at each integer time. The black lines represent theoretical predication. The dashed lines represent critical time $\tau_{crt}$.

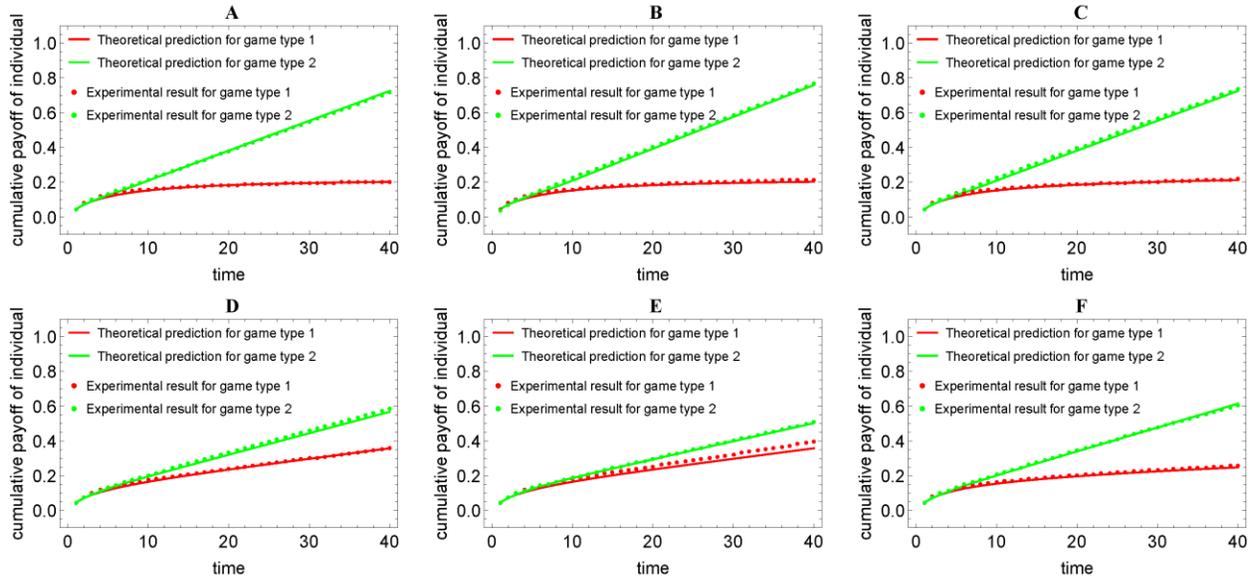

**Fig. S25**. **The comparison of average individual cumulative payoff for both game types between experimental result and theoretical prediction.** Parameters in these games are size $N = 20$, growth rate $T = 2$, normalized extraction parameter $\hat{e}_C = 0.7, \hat{e}_D = 1.1$, initial condition $R_0 = 0.5, x_0 = 0.5$ and game time $t_f = 40$. (**A**) complete network run in China; (**B**) BA network run in China; (**C**) SW network run in China; (**D**) complete network run in USA; (**E**) complete network run in USA; (**F**) complete network run in USA. Red and green points represent average all realizations of game type 1 and 2 at each integer time. Red and green line represent their theoretical prediction.

**Table S1. Recorded entities in the dataset.**

| Entity name | Meaning |
| --- | --- |
| Index | Node index of player who makes this decision |
| strategy | Strategy of player's present decision |
| last_strategy | Strategy of player's last decision |
| last_income | Payoff of player's last decision |
| neighbor_index | Strategy of last decision of one of player's neighbor |
| neighbor_income | Payoff of last decision of one of player's neighbor |
| R | Last resource volume |
| x | Last cooperation fraction |
| vector | Strategy vector of all players after player's decision |
| time | Time of decision |